\documentclass[twoside, twocolumn, 9pt]{article}
\usepackage[utf8]{inputenc}
\usepackage[T1]{fontenc}
\usepackage{graphicx}
\usepackage{longtable}
\usepackage{wrapfig}
\usepackage{rotating}
\usepackage[normalem]{ulem}
\usepackage{amsmath}
\usepackage{amssymb}
\usepackage{capt-of}
\usepackage{hyperref}
\usepackage[acronym,nogroupskip,nonumberlist]{glossaries}
\usepackage[stylemods]{glossaries-extra}
\makeglossaries
\usepackage{abstract}
\usepackage[export]{adjustbox}
\usepackage{sectsty}
\usepackage{setspace}
\usepackage{amsthm}
\usepackage{mathtools}
\usepackage{lastpage}
\usepackage{amsfonts}
\usepackage{mathptmx}
\usepackage{charter}
\usepackage{times}
\usepackage{bm}
\usepackage[
format=plain,
justification=justified,
singlelinecheck=false,
font={small, sf},
labelfont=bf,
labelsep=space
]{caption}

\usepackage[compact]{titlesec}
\usepackage{fancyhdr}
\usepackage{fnpos}
\usepackage{sidecap}
\usepackage{gensymb}
\usepackage{upgreek}
\usepackage{soul}

\usepackage{array}
\usepackage{droidsans}

\usepackage[usenames,dvipsnames]{xcolor}
\usepackage[version=3]{mhchem}
\usepackage[super,sort&compress,comma]{natbib}

\usepackage[
left=1.5cm,
right=1.5cm,
top=1.785cm,
bottom=2.0cm
]{geometry}

\definecolor{cream}{RGB}{222,217,201}

\newacronym{dft}{DFT}{density functional theory}
\newacronym{sqs}{SQS}{special quasi-random structures}
\newacronym{paw}{PAW}{projector augmented wave}
\newacronym{gga}{GGA}{generalized gradient approximation}
\newacronym{pbe}{PBE}{Perdew-Burke-Ernzerhof}
\newacronym{hse}{HSE}{Heyd-Scuseria-Ernzerhof}
\newacronym{soc}{SOC}{Spin Orbit Coupling}
\newacronym{ma}{MA}{Methylammonium}
\newacronym{fa}{FA}{Formamidinium}
\newacronym{pce}{PCE}{power conversion efficiency}
\newacronym{hap}{HaPs}{halide perovskites}
\newacronym{pca}{PCA}{principal component analysis}
\newacronym{tsne}{t-SNE}{t-distributed stochastic neighbor embedding}
\newacronym{umap}{UMAP}{uniform manifold approximation and projection}
\newglossaryentry{vasp}{name=VASP,description={{Vienna Ab initio Simulation Package}}}
\newglossaryentry{slme}{name=SLME,description={{spectroscopic limited maximum efficiency}}}
\newglossaryentry{qmml}{name=QM/ML,description={{quantum mechanics machine learning}}}
\date{\today}
\title{A High-Throughput Computational Dataset of Halide Perovskite Alloys\textsuperscript{\dag}}
\hypersetup{
 pdfauthor={Panayotis Manganaris},
 pdftitle={A High-Throughput Computational Dataset of Halide Perovskite Alloys\textsuperscript{\dag}},
 pdfkeywords={},
 pdfsubject={},
 pdfcreator={Emacs 29.0.50 (Org mode 9.5.3)}, 
 pdflang={English}}
\begin{document}


\pagestyle{fancy}
\thispagestyle{plain}
\fancypagestyle{plain}{
\renewcommand{\headrulewidth}{0pt}
}

\makeFNbottom
\makeatletter
\renewcommand\LARGE{\@setfontsize\LARGE{15pt}{17}}
\renewcommand\Large{\@setfontsize\Large{12pt}{14}}
\renewcommand\large{\@setfontsize\large{10pt}{12}}
\renewcommand\footnotesize{\@setfontsize\footnotesize{7pt}{10}}
\makeatother

\renewcommand{\thefootnote}{\fnsymbol{footnote}}
\renewcommand\footnoterule{\vspace*{1pt}%
\color{cream}\hrule width 3.5in height 0.4pt \color{black}\vspace*{5pt}} 
\setcounter{secnumdepth}{5}

\makeatletter 
\renewcommand\@biblabel[1]{#1}            
\renewcommand\@makefntext[1]{\noindent\makebox[0pt][r]{\@thefnmark\,}#1}
\makeatother 
\renewcommand{\figurename}{\small{Fig.}~}
\sectionfont{\sffamily\Large}
\subsectionfont{\normalsize}
\subsubsectionfont{\bf}
\setstretch{1.125}
\setlength{\skip\footins}{0.8cm}
\setlength{\footnotesep}{0.25cm}
\setlength{\jot}{10pt}
\titlespacing*{\section}{0pt}{4pt}{4pt}
\titlespacing*{\subsection}{0pt}{15pt}{1pt}

\fancyfoot{}
\fancyfoot[RO]{\footnotesize{\sffamily{1--\pageref{LastPage} ~\textbar  \hspace{2pt}\thepage}}}
\fancyfoot[LE]{\footnotesize{\sffamily{\thepage~\textbar\hspace{3.45cm} 1--\pageref{LastPage}}}}
\fancyhead{}
\renewcommand{\headrulewidth}{0pt} 
\renewcommand{\footrulewidth}{0pt}
\setlength{\arrayrulewidth}{1pt}
\setlength{\columnsep}{6.5mm}
\setlength\bibsep{1pt}

\makeatletter 
\newlength{\figrulesep} 
\setlength{\figrulesep}{0.5\textfloatsep} 

\newcommand{\topfigrule}{\vspace*{-1pt}%
\noindent{\color{cream}\rule[-\figrulesep]{\columnwidth}{1.5pt}} }

\newcommand{\botfigrule}{\vspace*{-2pt}%
\noindent{\color{cream}\rule[\figrulesep]{\columnwidth}{1.5pt}} }

\newcommand{\dblfigrule}{\vspace*{-1pt}%
\noindent{\color{cream}\rule[-\figrulesep]{\textwidth}{1.5pt}} }

\makeatother

\twocolumn[
\begin{@twocolumnfalse}
\vspace{1em}
\sffamily
\begin{tabular}{m{4.5cm} p{13.5cm} }
& \noindent\LARGE{\textbf{A High-Throughput Computational Dataset of Halide Perovskite Alloys\textsuperscript{\dag}}}\\%
\vspace{0.3cm} & \vspace{0.3cm} \\
& \noindent\large{Jiaqi Yang\textsuperscript{a}, Panayotis Manganaris\textsuperscript{a}, and Arun Mannodi-Kanakkithodi\textsuperscript{a}}\\
\end{tabular}

\begin{abstract}
Novel halide perovskites with improved stability and optoelectronic properties can be designed via composition engineering at cation and/or anion sites. Data-driven methods, especially involving high-throughput first principles computations and subsequent analysis based on unique materials descriptors, are key to achieving this goal. In this work, we report a density functional theory (DFT) based dataset of 495 ABX\textsubscript{3} halide
perovskite compounds, with various atomic and molecular species considered at A, B and X sites, and different amounts of mixing applied at each site using the special quasirandom structures (SQS) approach for alloys. We perform GGA-PBE calculations on all 495 pseudo-cubic perovskite structures and between 250 and 300 calculations each
using the more expensive HSE06 functional, with and without spin-orbit coupling, both including full geometry optimization and static calculations on PBE optimized structures. Lattice constants, decomposition energy, band gap, and theoretical photovoltaic efficiency derived from computed optical absorption spectra, are computed using each level of theory, and some comparisons are made with collected experimental values. Trends in the data are unraveled in terms of the effects of mixing at different sites, fractions of specific elemental or molecular species present in the compound, and averaged physical properties of species at different sites. We perform screening across the perovskite dataset based on multiple definitions of tolerance factors, deviation from cubicity in the optimization cell, and computed stability and optoelectronic properties, leading to a list of promising compositions as well as design principles for achieving multiple desired properties. Our multi-objective, multi-fidelity, computational halide perovskite alloy dataset, one of the most comprehensive to date, is available open-source, and currently being used to train predictive and optimization models for accelerating the design of novel compositions for superior performance across many optoelectronic applications.
\end{abstract}
\end{@twocolumnfalse}
\vspace{0.6cm}
]

\renewcommand*\rmdefault{bch}\normalfont\upshape
\rmfamily
\section*{}
\vspace{-1cm}


\footnotetext{%
  \textsuperscript{a}School of Materials Engineering, Purdue University, West Lafayette, IN 47907, USA; E-mail: amannodi@purdue.edu
}
\footnotetext{%
  \textsuperscript{\dag}Electronic Supplementary Information (ESI) available:
  \url{https://github.com/PanayotisManganaris/manuscript--PIP-data-manifest}
  See DOI: 00.0000/00000000.
}

\section{Introduction}

Perovskites have historically been materials of immense interest for a variety of industrial applications. With a general formula of ABX$_{3}$, a perovskite cubic unit cell contains two cations A and B at the corners and body center, and an anion X at each of the face centers. The symbolic 3D perovskite structure is a network of B$X_6$ octahedra robustly held together by large A-site cations. This unique structure means that perovskite properties are incredibly tunable, by changing the size and number of A/B/X species, by manipulating relative octahedral arrangements, and by creating non-cubic and metastable phases. Numerous research efforts have been devoted to \acrfull{hap}, especially as photovoltaic (PV) absorbers \cite{ansari-2018-front-oppor,yin-2015-halid-perov,manser-2016-intrig-optoel,brenner-2016-hybrid-organ}. In AB$X_3$
HaPs, X-site anions are halogens such as I and Br,
B-site cations may be divalent elements such as Pb and Sn, and the A-site is occupied by large monovalent cations that are either inorganic (e.g Cs, K, Rb) elements or organic
molecules (e.g \acrfull{ma} and \acrfull{fa}). The most
commonly studied hybrid organic-inorganic \acrlong{hap}, MAPbI$_{3}$ and FAPbI$_{3}$, have demonstrated large \acrfull{pce} values between 20\% and 25\% when used as absorbers in single- or multi-junction solar cells \cite{cui-2019-planar-p,jeong-2020-stabl-perov}. This is a five-fold improvement over the efficiencies first reported in 2009 and shows the most attractive feature of HaPs, their unique tunability. A perovskite structure is considered stable if the ionic radii of A, B, and X-site species satisfy the well-known tolerance (t) and octahedral (\(\mu\)) factors \cite{bartel-2019-new-toler}. Even under these restrictions, the chemical space of HaP structures, alloying ratios, ionic ordering, and possible defects, is still combinatorial and poses a highly multidimensional optimization problem.

Three of the most common ways of tuning the properties of HaPs are described below:

\begin{enumerate}

\item Composition: The most promising HaP compositions for PV absorption explored to date usually contain a mix of MA, FA, and Cs at the A-site, primarily Pb at the B-site with minor fractions of other divalent cations such as Sn and Ge, and I or Br at the X-site often with little Cl. Discovery of novel HaP compositions with attractive properties is on the rise as researchers expand the search into more complex alloys, novel A-site organic molecules, and substitutes for Pb at the B-site from Group IV, Group II, or transition elements. \cite{zhu-2019-struc-elect,banerjee-2019-rashb-trigg,ding-2019-cesium-decreas,greenland-2020-correl-phase}
Mixing at A site improves the general stability to degradation, while B site and X site mixing can tune and optimize band gaps and optical absorption. The allure of A/B/X-site mixing, even creating high entropy perovskite alloys, is in obtaining starkly different properties than pure compositions, possibly eliminating the harmful effect of defects, and improving the long-term stability and consequent optoelectronic performance.

\item Structure and phases: While the canonical perovskite phase is cubic, many HaPs are most stable in tetragonal,
orthorhombic, or hexagonal phases \cite{kar-2018-comput-screen}. For a given composition and phase, there may exist many local minima on the potential energy surface, typically sampled via rigorous application of evolutionary or minima hopping algorithms, atomic perturbations within larger supercells, or by varying degrees of distortion and rotation in the octahedral networks. Stable or metastable structures thus obtained may show better properties than previously studied ground state structures \cite{kim-2017-hybrid-organ}. In addition, HaPs may also manifest as double perovskites or 2D layered perovskites which include large organic spacer ligands, providing another means to tuning the stability and optoelectronic properties.

\item Defects: Investigation of the electronic structure of crystalline materials is incomplete without consideration of point defects, either native or impurity, which will affect the optoelectronic properties by modifying charge carrier lifetimes, equilibrium conductivity, and resulting trap-limited efficiencies. \cite{intro_defect_Dahliah,intro_defect_Kim} Point defects manifest as vacancies, interstitials, or substitutions, and the same defect may behave very differently in different compositions or structures- highlighting the need to include the presence of defects as another variable towards tuning HaP properties.

\end{enumerate}

The chemical design space of HaPs is very much combinatorial and raises challenges for experimentalists to perform effective screening. First principles-based density functional theory (DFT) simulations have been systematically performed to study the optoelectronic properties of HaPs as a function of structure, composition, and defects. The expense of standard DFT computations is reasonable when searching for new promising candidates in such a boundless space. Recently, DFT simulations have been reliably used for modeling structural information, heat of formation or decomposition, band gaps, optical absorption spectra,
and defect formation energies of a variety of HaPs. \cite{mannodi-kanakkithodi-2022-data-driven,yin-2015-halid-perov} High-throughput DFT (HT-DFT) computations provide the most effective way to screen across a large space of hybrid and inorganic ABX$_{3}$ halide perovskites. An examination of the HaP-related computational literature reveals that there have been numerous medium ($\sim$ 10$^2$ data points) to large ($\sim$ 10$^3$ or more data points) DFT datasets reported for HaPs, which have been successfully used to screen promising materials with desired stability and formability as well as PV-suitable band gaps, among other properties. \cite{castelli-2014-bandg-calcul,park-2019-explor-new,kar-2018-comput-screen,Pu_screening}

A clear limitation of HT-DFT-driven screening is the computational expense of applying a suitably advanced level of theory across a large number of materials. This problem is typically addressed by coupling DFT computations with state-of-the-art machine learning (ML) or artificial intelligence (AI) techniques. Within the area of perovskites, there are many examples in the literature where DFT datasets and suitable atomic/structural/compositional descriptors have been used to train a variety ML-based predictive and classification models, leading to accelerated prediction of lattice constants, formation energies, band gaps, and other important properties. \cite{park-2019-explor-new,JaredStanley_ML_LeadFree,LEE_HOIP} Such DFT-ML models, once rigorously trained and tested, are deployed for high-throughput screening across massive datasets of unknown perovskites. We recently published a systematic overview of many such efforts applying DFT and/or ML towards halide perovskite discovery. \cite{MRS_Bulletin_JY_Arun}

In this work, we report a large HT-DFT dataset of 495 chemically distinct, pseudo-cubic, halide perovskite alloys. This dataset builds upon the 229 compounds reported in prior work by \citet{mannodi-kanakkithodi-2022-data-driven}, adding more types of mixing, better property estimates, and thorough analysis of trends and correlations. The relatively large size of this dataset is intended to provide an initial sampling suitable for a guided search within the HaP alloy space. In this dataset, all perovskite structures are cubic or pseudo-cubic, and the focus is more on investigating the dependence of computed properties on composition, and specifically the type of alloying.

\begin{figure*}[t]
\centering
\includegraphics[width=.9\linewidth]{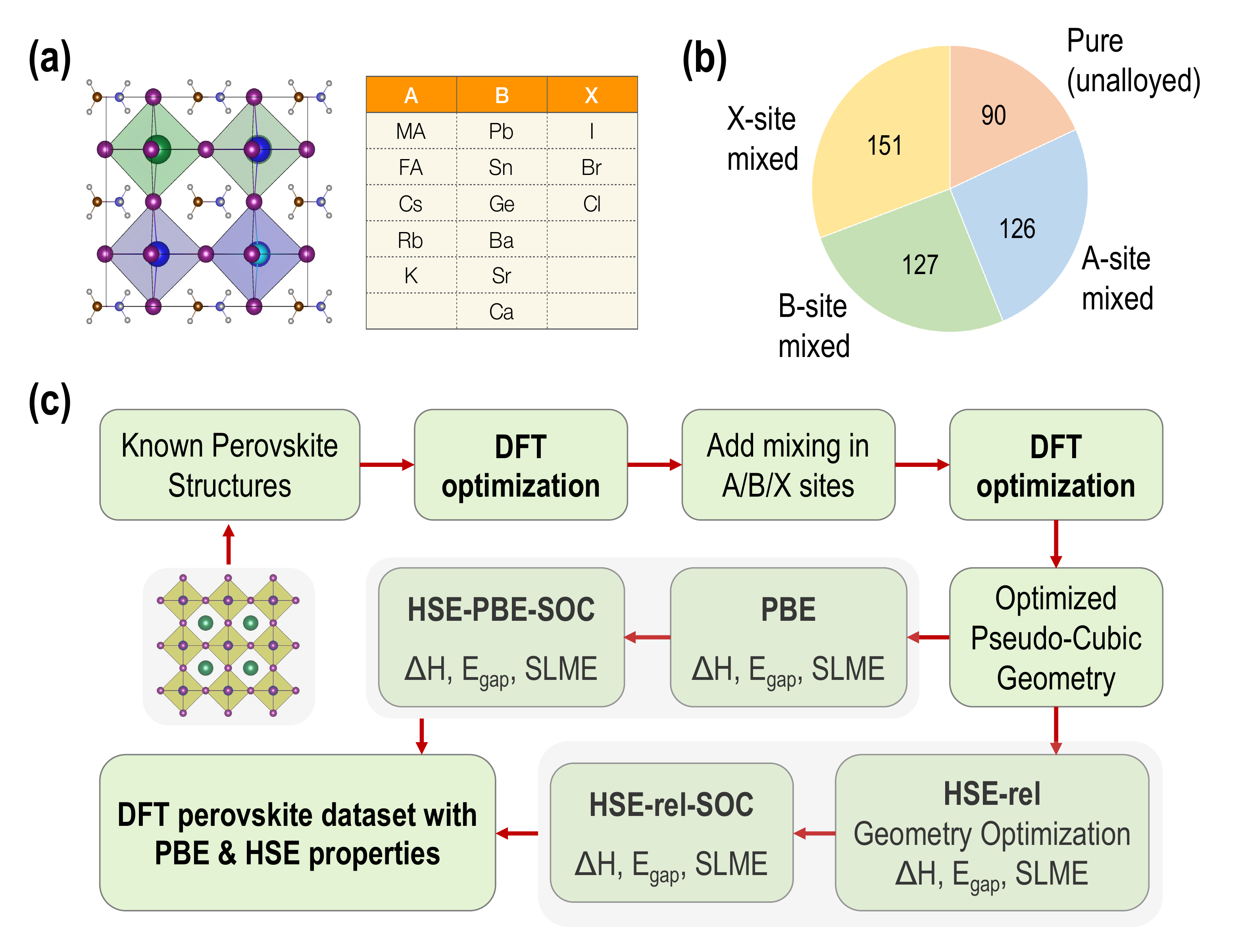}
\caption{\label{fig:outline} (a) Chemical space of ABX\textsubscript{3} perovskites studied in this work. (b) Number of samples representing each kind of primary alloy. (c) Steps involved in generating the PBE and HSE datasets of three kinds of properties, namely decomposition energy ($\Delta$H), band gap (E$_{gap}$), and spectroscopic limited maximum efficiency (SLME).}
\end{figure*}

Based on the generated perovskite structures, we perform GGA-PBE calculations and report the computed decomposition energy, band gap, and theoretical PV efficiency. In addition, around 250 to 300 calculations are performed using the HSE06 functional (henceforth referred to as HSE), in three different flavors: using full geometry optimization, with and without spin-orbit coupling (SOC), and static calculations on GGA-optimized structures with SOC. The same properties are computed from all three types of HSE computations: this enables the comparison of PBE and multiple HSE estimates with experiments, as well as an understanding of the importance of SOC for certain compositions. Pearson correlation analysis is performed to study the contribution of specific A/B/X species and their known elemental/molecular properties on the DFT computed properties, leading to some useful design rules. We further combine DFT-computed properties with perovskite stability factors such as the octahedral and tolerance factors, and determine a deviation from cubicity for all optimized structures, to obtain a list of promising candidates for solar absorption and related optoelectronic applications. We emphasize that this chemically diverse, multi-objective, multi-fidelity dataset of HaP alloys will serve many ML endeavors in the future for prediction and inverse design, be used as the foundation for extended datasets of non-cubic structures and new chemistries, and drive the experimental discovery of novel HaP compositions with targeted properties.

\section{Methodology}

\subsection{Devising a Halide Perovskite Chemical Space}

The dataset we report is based on the standard cubic ABX\textsubscript{3} perovskite structure. Fourteen common perovskite constituents are selected to form the chemical space. The five constituents making up the A-site occupants include three inorganic and two organic cations. Six divalent metals represent the possible B-site occupants and three halogen anions make up the possible X-site occupants. The elemental and molecular space used to construct the data set is shown in Fig. \ref{fig:outline}(a). In total, these component vectors form a constrained 14 dimensional space within which all perovskite compounds consisting of the species shown in Fig. \ref{fig:outline} (a) must exist.

\begin{figure*}[t]
\centering
\includegraphics[width=1.0\linewidth]{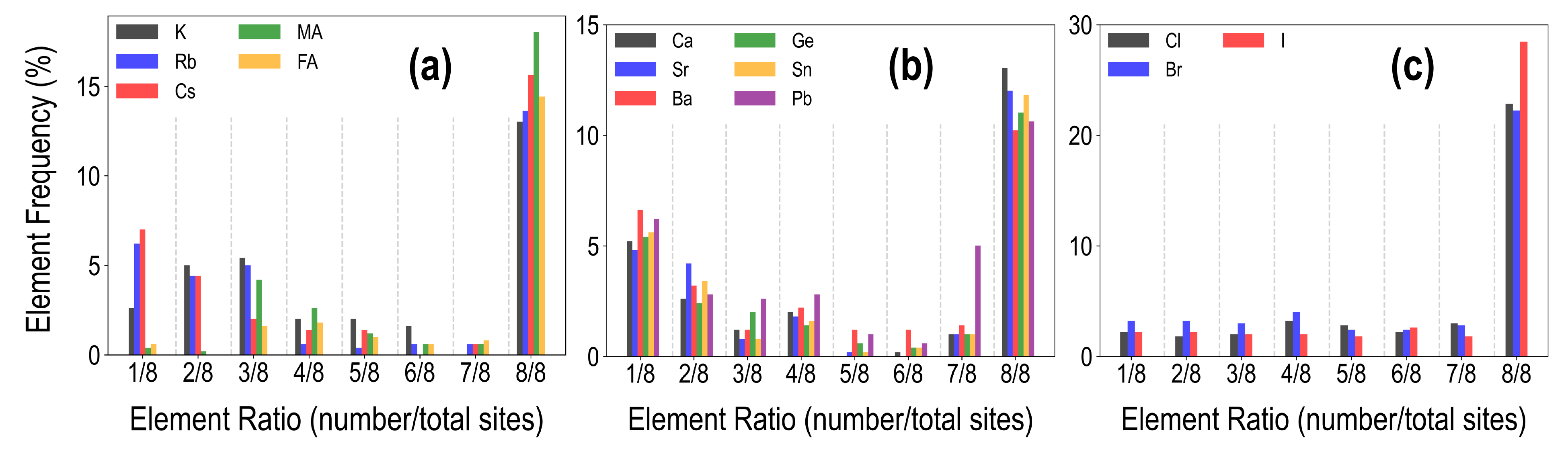}
\caption{\label{fig:PBE_freq} Distribution of mixing fractions of various species at the A (a), B (b), and X (c) sites across the PBE dataset of 495 compounds.}
\end{figure*}

The pure (non-alloyed) possibilities are exhaustively sampled using \(5*6*3 = 90\) compounds. Starting from these pure perovskite structures, we perform systematic mixing at the A, B, and X sites. For simulating perovskite alloys, the \acrfull{sqs} method \cite{jiang-2016-special-quasir} is applied to build periodic structures that make the first nearest-neighbor shells as similar to the target random alloy as possible. The \acrshort{sqs} can be considered the best possible periodic supercell representing a given mixed-composition perovskite. The distribution of different types of mixing across our dataset  is shown in Fig. \ref{fig:outline}(b). For simplicity, only one type of mixing at a time is considered in this study; that is, mixing is not performed at multiple (A/B/X) sites simultaneously. In total, we performed GGA-PBE computations on 90 pure, 126 A-site mixed, 151 B-site mixed, and 127 X-site mixed HaPs.

Each HaP composition is simulated using a 2x2x2 supercell, which allows A- and B-site mixing to be performed in discrete 1/8\textsuperscript{th} fractions of the total site occupancy, and X-site mixing in 1/24\textsuperscript{th} fractions, though for simplicity, we restrict X-site mixing to fractions of 3x/24. At these mixing levels, it is appropriate to call all of these perovskites alloys. Fig. \ref{fig:PBE_freq} shows the distribution of various types of mixing of the 14 elemental and molecular species at the A, B, and X-sites, across the dataset of 495 compounds. Since mixing is only allowed on one out of the three sites at a time, there is a higher prevalence of the 8/8 fraction for each species. We also find a larger occurrence of the smallest fractions of mixing at A and B sites as compared to intermediate fractions; overall, every type of mixing is represented within the dataset a few times. Using the procedure presented in Fig. \ref{fig:outline}(c), we calculate the stability and optoelectronic properties for the HaP dataset using both the semilocal GGA-PBE functional and the hybrid HSE06 functionals. Ultimately, we generated a dataset of 495 points at the PBE level, and ~300 points each at the HSE-PBE+SOC (refers to HSE+SOC on PBE relaxed structures), HSE-relaxed, and HSE-relaxed+SOC levels. The exact constitution of the dataset from different levels of theory is presented in Table \ref{table:data_num}.

\begin{table}[!ht]
    \centering
    \begin{tabular}{|c|c|}
    \hline
        \textbf{Functional} & \textbf{Number of Data Points} \\ \hline
        PBE & 495 \\ \hline
        HSE-rel & 299 \\ \hline
        HSE-rel-SOC & 282 \\ \hline
        HSE-PBE-SOC & 244 \\ \hline
    \end{tabular}
    \caption{\label{table:data_num} Number of HaP compounds studied using each of the 4 theories applied in this work.}
\end{table}

\subsection{DFT Details}

All DFT computations were performed using \gls{vasp} version 6.2 \cite{vasp1,vasp2,vasp3} employing the \acrfull{paw} pseudopotentials. \cite{PAW1,PAW2} The \acrfull{pbe} functional within the \acrfull{gga} \cite{vasp_pbe} as well as the hybrid \acrshort{hse}06\cite{HSE06} (\(\alpha\)=0.25 and \(\omega\)=0.2) functionals are used for exchange-correlation energy. The energy cutoff for the plane-wave basis is set to 500 eV. For all \acrshort{pbe} geometry optimization calculations, the Brillouin zone was sampled using a 6$\times$6$\times$6 Monkhorst-Pack mesh for unit cells and a 3$\times$3$\times$3 for supercells. Using the PBE optimized structure as input, the electronic band structure is calculated along high-symmetry k-points \cite{Band_structure,sumo} to obtain accurate band gaps, and the optical absorption spectrum is further calculated using the LOPTICS tag, setting the number of energy bands to 1000 for each structure. For \acrshort{hse} calculations, geometry optimization was performed using only the Gamma point, and subsequent computations used a reduced 2$\times$2$\times$2 Monkhorst-Pack mesh. The force convergence threshold is set to be -0.05 eV/\AA{}. Spin-orbit coupling is also applied to two flavors of HSE computations using the LORBIT tag and the non-collinear magnetic version of VASP 6.2. \cite{VASP_SOC} We obtain optical absorption spectra from different HSE functionals by using the difference between the respective PBE and HSE band gap, and shifting the PBE-computed spectrum.

\subsection{DFT Computed Properties}

\subsubsection{Decomposition Energy}

In this work, we estimate the stability of any ABX$_{3}$ compound based on the energy of decomposition to all possible AX and B$X_2$ phases. In addition, we add a mixing entropy term for all alloys, assuming the temperature to be 300K. The decomposition energy ($\Delta$H) is thus calculated using equation \eqref{eq:decoE}, individually from every level of theory.

\begin{equation}\label{eq:decoE}
\begin{aligned}
    \Delta H = E_{opt}(ABX_3) - \sum_{i} x_{i}E_{opt}(AX) - \sum_{i} x_{i}E_{opt}(BX_2) \\ 
    + k_{B}T(\sum_{i} x_{i}ln(x_{i}))
\end{aligned}
\end{equation}

\begin{figure*}[t]
\centering
\includegraphics[width=1.0\linewidth]{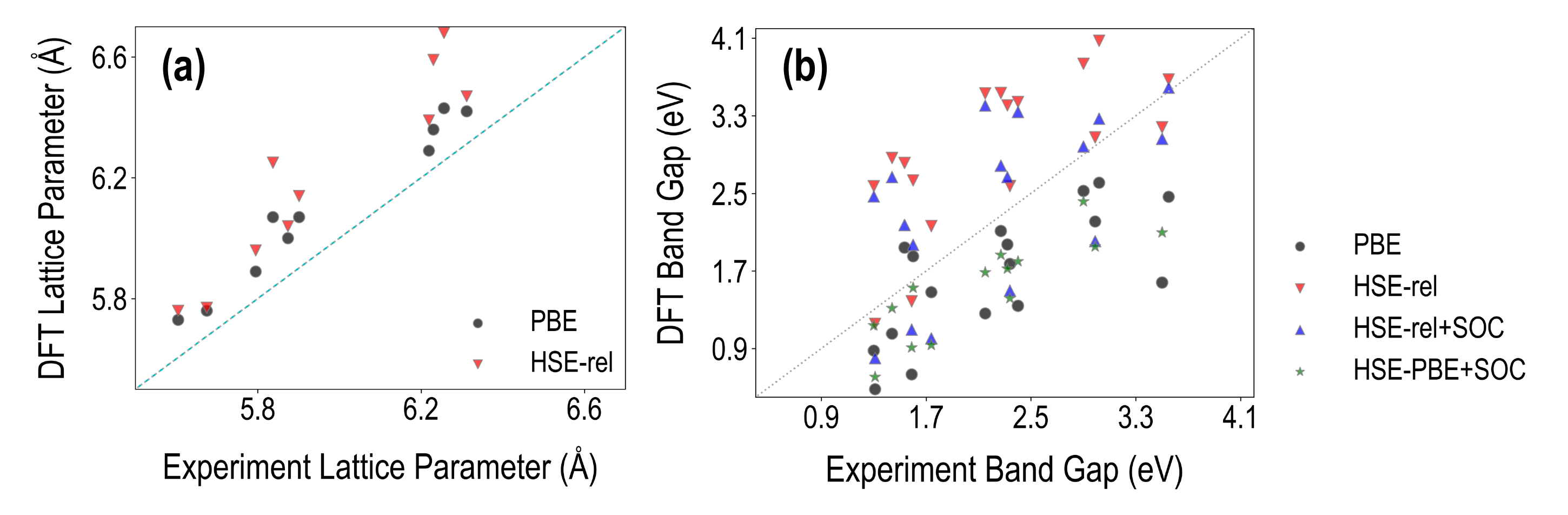}
\caption{\label{fig:comp_pbe_exp} Comparison between DFT computed and experimentally measured properties of selected HaPs: (a) cubic lattice constants, and (b) band gaps.}
\end{figure*}

Here, E$_{opt}$(M) refers to the total DFT energy of any compound M, k$_{B}$ is the Boltzmann constant, T is the temperature (fixed to be 300K in this work), and x$_{i}$ is the fraction of any particular species mixed at the A/B/X site. The weighted sums over E$_{opt}$(AX) and E$_{opt}$(BX$_2$) signify that an ABX$_3$ alloy is assumed to decompose to multiple AX and BX$_2$ phases, based on the number of species mixed at the A, B, or X site. Taking A$(B_1)_x(B_2)_{1-x}X_3$ as an example, the decomposition energy would be calculated using equation \eqref{eq:decoE_2}. We assume that the ``BX$_2$" decomposition products for B1-B2 mixed perovskite are ($B_1$)$X_2$ and ($B_2$)$X_2$.

\begin{equation}\label{eq:decoE_2}
\begin{aligned}
\Delta H[A(B_1)_x(B_2)_{1-x}X_3] = E_{opt}(AB_1B_2X_3) \\
- E_{opt}(AX) - x*E_{opt}(B_1X_2)-(1-x)*E_{opt}(B_2X_2) \\
+ k_{B}T(xln(x) + (1-x)ln(1-x))
\end{aligned}
\end{equation}

The decomposition energy will be calculated from 4 different levels of theory, namely PBE ($\Delta$H$^{PBE}$), HSE-relaxed ($\Delta$H$^{HSE-rel}$), HSE-relaxed-SOC ($\Delta$H$^{HSE-rel-SOC}$), and HSE-PBE-SOC ($\Delta$H$^{HSE-PBE-SOC}$). All decomposition energy values are reported per ABX$_3$ formula unit. Calculating $\Delta$H for X-site mixed compounds involves some additional work because of the multiple choices for AX and BX$_2$ phases; more details are provided in the SI and in Figs. S1 and S2.

\subsubsection{Band Gap}

From the PBE band structure calculations and the static HSE calculations using the 2$\times$2$\times$2 Monkhorst-Pack mesh, four types of electronic band gaps are computed in eV: PBE (E$_{gap}$$^{PBE}$), HSE-relaxed (E$_{gap}$$^{HSE-rel}$), HSE-relaxed-SOC (E$_{gap}$$^{HSE-rel-SOC}$), and HSE-PBE-SOC (E$_{gap}$$^{HSE-PBE-SOC}$).

\subsubsection{Spectroscopic Limited Maximum Efficiency (SLME)}
\label{sec:org21fb725}
Introduced by \citet{yu-2012-ident-poten}, the \gls{slme} is a convenient metric for evaluating a semiconductor's suitability for single junction photovoltaic (PV) absorption. In this work, \gls{slme} is calculated considering a 5\(\mu\)m sample thickness for every perovskite using equation \eqref{eq:absorption_alpha},  \eqref{eq:slme_int}, and \eqref{eq:slme_sum}

\begin{equation}\label{eq:absorption_alpha}
a(E)=1-e^{-2\alpha(E)L}
\end{equation}

Here, \(\alpha(E)\) is the DFT computed optical absorption coefficient as a function of incident photon energy and \(L\) is the thickness of the absorber.

\begin{equation}\label{eq:slme_int}
J=e\int_{0}^{\infty} a(E)I_{sun}(E)dE - J_{0}(1-e^{\frac{eV}{kT}})
\end{equation}

\begin{equation}\label{eq:slme_sum}
\eta = \frac{P_{m}}{P_{in}}=\frac{max(J \times V)}{P_{in}}
\end{equation}

J is the current density, $I_{sun}$ is the light spectrum intensity of sunlight, and P refers to the power used to calculate SLME efficiency. Using the DFT computed optical absorption spectrum as well as the magnitude and type (direct of indirect) of band gap as input, SLME is directly calculated using an open-source package \cite{williams-2022-sl3me}. This calculation is performed using PBE as well as the 3 different HSE functionals, resulting in 4 theoretical estimates of PV efficiency, denoted as SLME$^{PBE}$, SLME$^{HSE-rel}$, SLME$^{HSE-rel-SOC}$, and SLME$^{HSE-PBE-SOC}$.

\section{Results and discussion}

\subsection{Comparing DFT with Experiments}

\begin{table}[!ht]
    \centering
    \begin{tabular}{|c|c|}
    \hline
        \textbf{Functional} & \textbf{Band Gap RMSE vs Exp (eV)} \\ \hline
        PBE & 0.78  \\ \hline
        HSE-rel & 0.93  \\ \hline
        HSE-rel-SOC & 0.74  \\ \hline
        HSE-PBE-SOC & 0.70 \\ \hline
    \end{tabular}
    \caption{\label{table:rmse_exp} RMSE values of band gaps computed from different functionals compared with experimental (Exp) values.}
\end{table}

In Fig \ref{fig:comp_pbe_exp}, we compare the various PBE and HSE calculated lattice constant and band gap values with corresponding experimental results collected from Tao el al\cite{exp_Tao} and Almora et al\cite{exp_almora}. We find that the root mean square error (RMSE) of PBE lattice constants compared to experiments is 0.27 \AA, while the corresponding HSE RMSE is 0.31 \AA, showing that a hybrid functional-based geometry optimization is unnecessary for obtaining accurate crystal structure information. We note that accuracy of optimized geometry may be further improved by using the PBEsol functional \cite{PBEsol} or by incorporating van der Waals interactions with the PBE functional using DFT-D3 \cite{PBE-D3} or a similar approach, especially for hybrid perovskites.

Fig \ref{fig:comp_pbe_exp}(b) shows that E$_{gap}$$^{PBE}$ is generally an underestimation compared to experiment, as expected, showing an RMSE of 0.78 eV. The corresponding RMSEs of E$_{gap}$$^{HSE-rel}$, E$_{gap}$$^{HSE-rel-SOC}$, and E$_{gap}$$^{HSE-PBE-SOC}$ are respectively 0.93 eV, 0.74 eV, and 0.70 eV. We find that on average, HSE-PBE-SOC is the best approach out of the four for reproducing band gaps, but other functionals may be more accurate for certain types of compositions (such as purely inorganic vs organic-inorganic, Pb-based or Pb-free, etc.), as will be discussed further later in this article. HSE-relaxed band gaps are heavily overestimated and brought down by the inclusion of SOC. It should also be noted that phase information was not always available for certain experimental data points collected from the literature, and non-cubic phases may affect the accuracy of the computational results here. Finally, it should be noted here that the PBE RMSE is not significantly different from the RMSE of HSE-PBE-SOC, which comes from the accidental accuracy of semi-local functionals without SOC for hybrid organic-inorganic perovskites. \cite{mannodi-kanakkithodi-2019-compr-comput,mannodi-kanakkithodi-2022-data-driven}

\begin{figure*}[h]
\centering
\includegraphics[width=.9\linewidth]{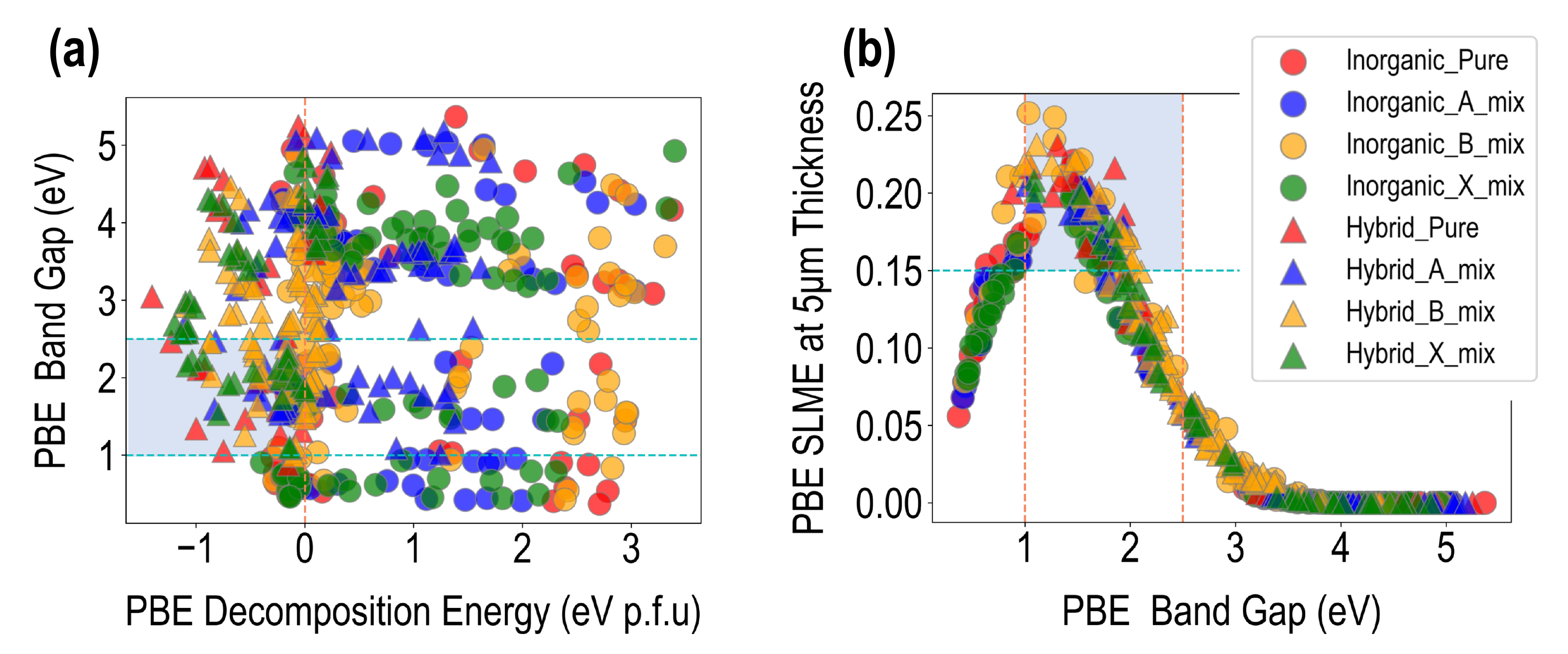}
\caption{\label{fig:pairplots} Visualization of the PBE dataset: (a) band gap against decomposition energy, and (b) SLME at 5$\mu$m thickness sample thickness against band gap. Different colors represent different types of mixing and different symbols are used to distinguish between purely inorganic and hybrid organic-inorganic HaPs.}
\end{figure*}

\subsection{Visualizing the PBE Dataset}

Fig \ref{fig:pairplots} presents a visualization of the PBE computed properties across the dataset of 495 compounds. The data can be distinguished in terms of purely inorganic vs hybrid organic-inorganic compounds, as well as in terms of the type of mixing. A broad range of values is observed for the three properties, which is a testament to the chemical diversity in our dataset. $\Delta$H$^{PBE}$ varies from $\sim$ -1.5 eV to $\sim$ 4 eV, with a majority of the data points in the unstable > 0 eV region, while E$_{gap}$$^{PBE}$ goes from $\sim$ 0.5 eV to $\sim$ 5.5 eV. SLME$^{PBE}$ goes from a low of 0 (when band gaps are too large for visible range absorption) to a maximum of 0.25 (or 25 \% efficiency). Fig \ref{fig:pairplots}(a) shows E$_{gap}$$^{PBE}$ plotted against $\Delta$H$^{PBE}$, with the shaded region showing the ranges of favorability, chosen here as $\Delta$H$^{PBE}$ < 0 eV and 1 eV < E$_{gap}$$^{PBE}$ < 2.5 eV. 

We find that stable compositions ($\Delta$H$^{PBE}$ < 0 eV) are predominantly occupied by hybrid organic-inorganic HaPs, with a fair few pure, B-site mixed, and X-site mixed compounds. A large number of A-site mixed hybrid HaPs as well as a majority of inorganic HaPs occupy the unstable region, indicating that although the presence of organic cations is desirable to prevent perovskite decomposition, mixing at the A-site may not always be beneficial. Band gap shows less clear trends, and as will be explained later, is largely dependent on the type and number of specific B and X-site ions. The region of desirable E$_{gap}$$^{PBE}$ and $\Delta$H$^{PBE}$ is largely populated by hybrid HaPs with B-site or X-site mixing. Furthermore, Fig \ref{fig:pairplots}(b) shows SLME$^{PBE}$ plotted against E$_{gap}$$^{PBE}$, showing the characteristic relationship that has been explored in past works. \cite{slme_1,slme_2} SLME rises initially as the band gap increases, reaching a peak of $\sim$ 25 \% around E$_{gap}$$^{PBE}$ = 1.5 eV, and subsequently goes down until it goes to 0 for E$_{gap}$$^{PBE}$ > $\sim$ 3 eV. The largest SLME$^{PBE}$ values are shown by pure hybrid and B-site mixed compounds, both hybrid and inorganic.

\subsection{Composition-Property Correlations}

\begin{figure*}[h]
\centering
\includegraphics[width=1.0\linewidth]{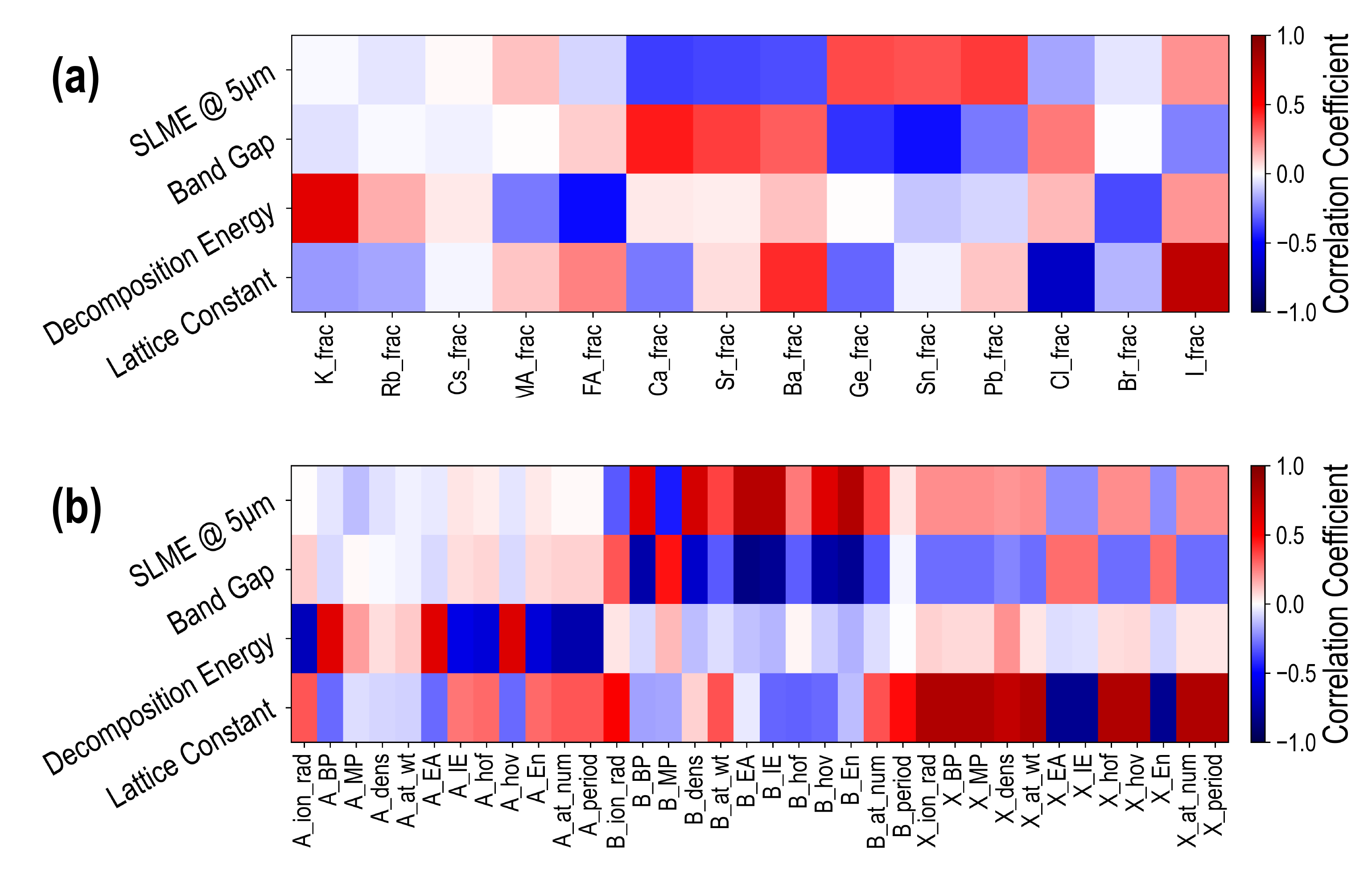}
\caption{\label{fig:pearson_pcomp} Pearson coefficients of linear correlation between 4 PBE computed properties and (a) 14 compositional descriptors, and (b) 36 elemental property descriptors.}
\end{figure*}

To obtain a qualitative understanding of how different constituents at the A, B, and X sites contribute to the properties of interest, we encode each compound in the dataset using a set of descriptors and calculate the Person coefficient of linear correlation \cite{PearsonCorr} between each descriptor dimension and each property. Since all HaPs in this study are cubic or pseudo-cubic, the essential distinguishing feature from one compound to another is the composition or the chemical formula. Every compound is thus encoded using two types of descriptors: a 14-dimensional composition vector representing fractions of every species (Cs, MA, Pb, Br, etc.) in the compound, and a 36-dimensional ``elemental properties" vector, representing weighted averages of 12 elemental properties each (such as ionic radii, electron affinity, ionization energy, etc.) of the respective species at A, B, and X sites. A complete list of all 50 descriptors is provided in Table SI.

Fig. \ref{fig:pearson_pcomp}(a) shows the linear correlation between composition descriptors and PBE properties, namely lattice constant, decomposition energy, band gap and SLME. In the heatmap, darker red implies large positive correlation, darker blue implies a large negative correlation, and white means there is very little or no correlation. A few important relationships immediately jump out from this plot: large ions like Ba and I lead to an increase in the lattice constant, while Cl has the reverse effect. An increase in the fraction of K at the A-site increases $\Delta$H$^{PBE}$ and thus makes the compound more unstable, while increasing the fraction of FA will make it more stable. B-site elements generally have little effect on the stability, but have much larger correlations with E$_{gap}$$^{PBE}$ and SLME$^{PBE}$. While Ca, Sr, and Ba increase E$_{gap}$$^{PBE}$ and decrease SLME$^{PBE}$, Ge and Sn decrease E$_{gap}$$^{PBE}$ and increase SLME$^{PBE}$. Pb shows a large positive correlation with SLME$^{PBE}$, which is not surprising given that FA/MA/Cs-based Pb iodide or bromide perovskites are most common in optoelectronic applications. Finally, X-site species show more modest correlation with E$_{gap}$$^{PBE}$ and SLME$^{PBE}$, with Br showing virtually no correlation with band gap, which is an effect of Br lying between I and Cl in the band gap spectrum. The lower values of correlation between X-site constituents and band gap and SLME reveals that mixing or complete substitution at the B-site has a more significant effect on the optoelectronic properties. These correlations provide confirmation for some well known effects and some simple design principles for HaP compositions with targeted properties.

Next, we calculated the linear correlations between the 36-dimensional elemental property descriptors and the 4 PBE properties, and the results are presented in Fig. \ref{fig:pearson_pcomp}(b). Once again, it can be seen that the biggest contributors to $\Delta$H$^{PBE}$ are A-site properties: specifically, increasing the ionic radius, ionization energy, or atomic number of A-site species makes the compound more stable, whereas increasing the boiling point, electron affinity, or heat of vaporization makes it less stable. The largest correlations with lattice constant are from X-site features, with higher electron affinity, ionization energy, or electronegativity of the X-site constituent reducing the lattice constant and all other properties increasing it. When it comes to band gap and SLME, we once again notice an overwhelming contribution from the B-site species. Increasing the boiling point, electron affinity, ionization energy, or electronegativity of B-site species helps decrease E$_{gap}$$^{PBE}$ as well as increase SLME$^{PBE}$, explaining why Pb/Sn/Ge are clearly more beneficial in PV applications that Ba/Sr/Ca at the B-site. These correlations help expand our design principles based purely on fractions of different species, and provide an opportunity to train predictive models for various properties. \cite{mannodi-kanakkithodi-2022-data-driven,MRS_Bulletin_JY_Arun}

\subsection{Improving Property Predictions using HSE06 and Spin-Orbit Coupling}

\begin{figure*}[h]
\centering
\includegraphics[width=1.0\linewidth]{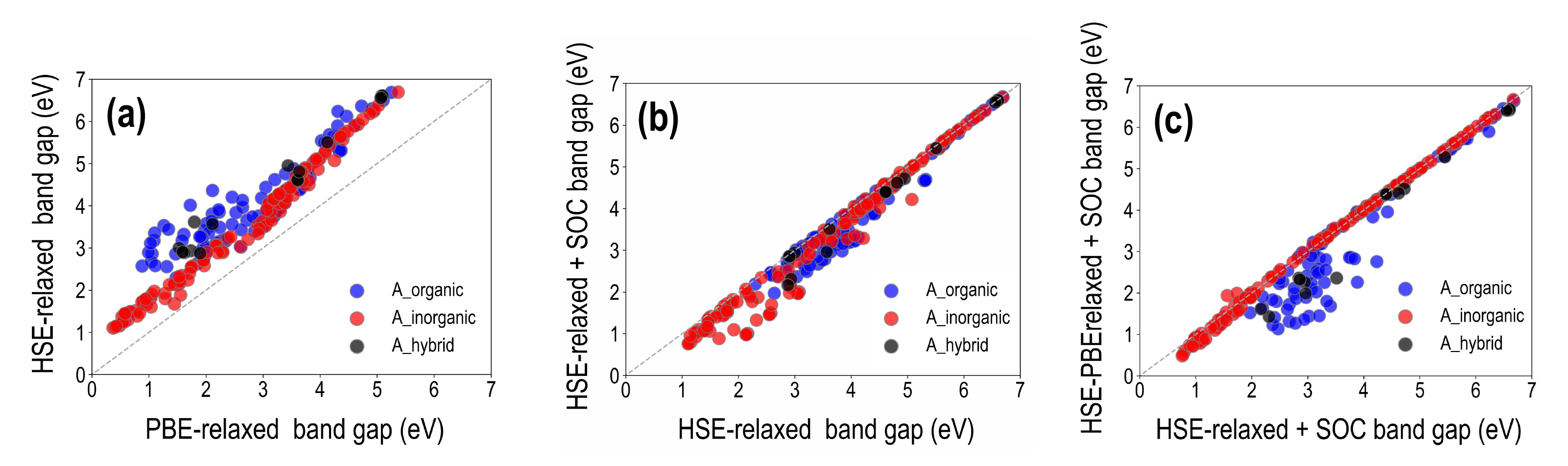}
\caption{\label{fig:sum_HSE_SOC} Visualization of band gaps computed from various HSE06 approaches: (a) HSE-relaxed band gap vs PBE band gap. (b) HSE-relaxed band gap with SOC vs HSE-relaxed band gap without SOC. (c) HSE-PBE with SOC band gap vs HSE-relaxed with SOC band gap.}
\end{figure*}

It was shown in section 3.1 that for a set of selected HaP compositions, while PBE-optimized lattice constants match well with experiments, PBE band gaps are underestimated, and HSE-PBE-SOC band gaps match better with measured values. GGA-PBE computations are generally reliable for structure and stability (formation or decomposition energy) of both hybrid and purely inorganic HaPs, but advanced levels of theory such as the HSE06 functional or GW approximation, with the inclusion of SOC to account for the relativisitic effects of heavy atoms such as Pb, are of paramount importance when it comes to electronic and optical properties. Here, we perform a series of expensive HSE calculations across the HaP dataset and report trends and major observations, specifically the effect of full geometry optimization with HSE compared to using the PBE-optimized structures, and the effect of incorporating SOC in the calculation. Overall, we generate HSE datasets of decomposition energy, band gap, and SLME, for HSE-relaxed (299 data points), HSE-relaxed+SOC (282 data points), and HSE-PBE-SOC (244 data points). 

A visualization of the types of mixing per A/B/X-site species is presented in Fig. S3, and different properties are plotted against each other for the three types of HSE datasets in Figs. S4, S5, and S6. We find similar distributions to the PBE data, with notable differences coming from HSE band gaps being generally larger and eliminating a lot of the low SLME data points. Very similar $\Delta$H values are obtained for all compositions from the 4 methods, showing that PBE-based stability metrics should be more than reliable. We note here that SLME from the different HSE functionals is obtained using the PBE-computed optical absorption spectrum shifted along the energy axis by the difference between E$_{gap}^{PBE}$ and the corresponding HSE E$_{gap}$: this is a method that helps us determine a theoretical efficiency from HSE without performing a full optical absorption calculation using HSE. Fig \ref{fig:sum_HSE_SOC} presents a comparison between the different types of HSE and PBE band gaps, dividing the data in terms of the nature of A-site species: purely organic, purely inorganic, or mixed organic-inorganic. While it is clear that B-site and X-site species are the major contributors to band gap, we divide the data like this mainly to observe how important HSE vs PBE geometry optimization is for hybrid vs inorganic HaPs, and the magnitudes of difference between PBE and HSE band gaps and between using and not using SOC. 

It can be seen from the 299 data points in Fig \ref{fig:sum_HSE_SOC}(a) that E$_{gap}$$^{HSE-rel}$ are, on average, 1 eV or more greater than E$_{gap}$$^{PBE}$, with larger differences appearing when A-site contains only organic molecules: we attribute this to the larger degree of geometry optimization from HSE in the presence of organic cations than when only inorganic cations are present, leading to larger differences in the band gap. Fig \ref{fig:sum_HSE_SOC}(b) shows E$_{gap}$$^{HSE-rel-SOC}$ plotted against E$_{gap}$$^{HSE-rel}$ for 282 data points. As expected, SOC brings down E$_{gap}$ for many of the compounds, and keeps E$_{gap}$ the same for many other compounds, confirming that SOC is certainly vital for certain compositions but can be ignored in others, as has been discussed in past studies. \cite{mannodi-kanakkithodi-2019-compr-comput,mannodi-kanakkithodi-2022-data-driven,T_Das} Interestingly, we observe that for several purely inorganic HaPs with lower E$_{gap}$ < 3 eV, SOC significantly reduces the gap. Next, we plot in Fig \ref{fig:sum_HSE_SOC}(c) E$_{gap}$$^{HSE-PBE-SOC}$ vs E$_{gap}$$^{HSE-rel-SOC}$ for 244 data points, in an attempt to understand the difference between HSE-relaxed and HSE-on-PBE-relaxed band gaps (with the inclusion of SOC in both). We find virtually identical band gaps from both methods for all pure inorganic HaPs, but large differences when organic cations exist at the A-site, which can once again be explained by the more severe geometry optimization from HSE in the latter.

\begin{figure}[h]
\centering
\includegraphics[width=1.0\linewidth]{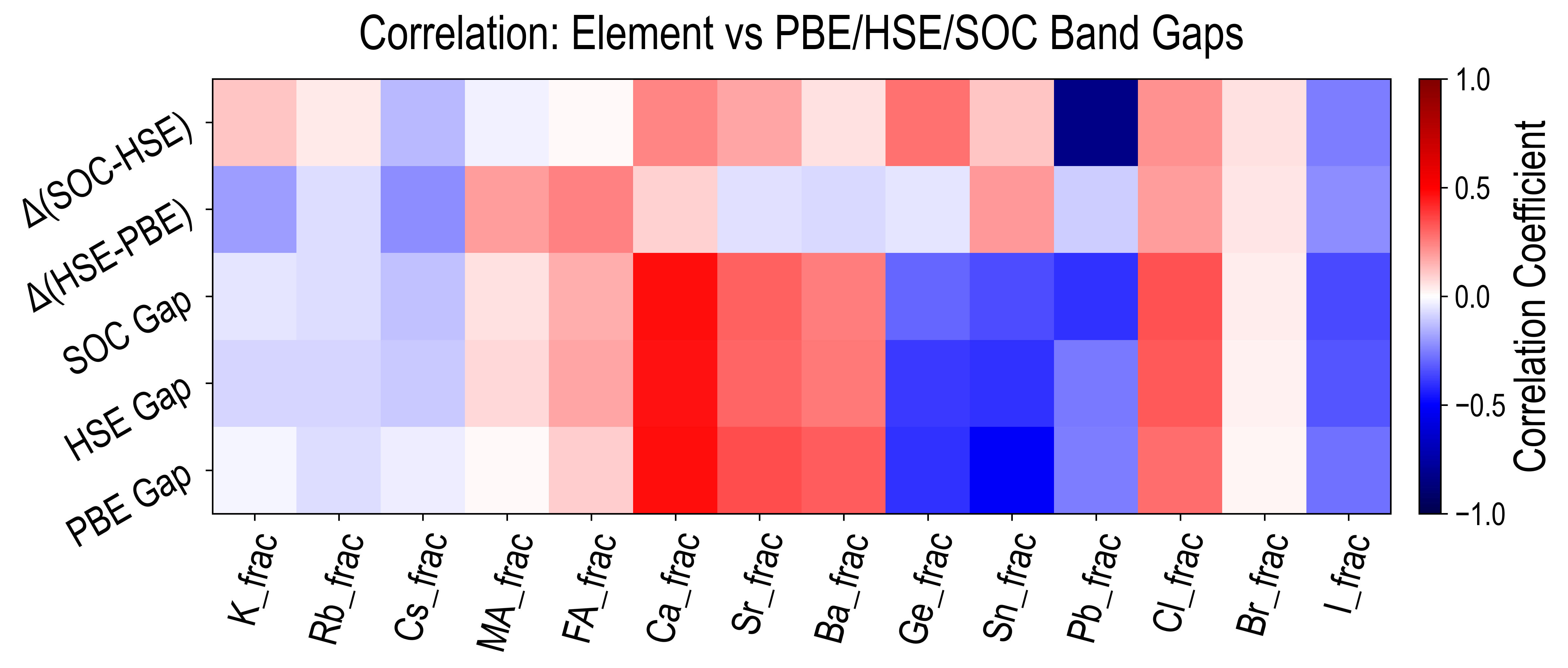}
\caption{\label{fig:pearson_HSE} Pearson coefficients of linear correlation between various types of band gaps and the 14 compositional descriptors. HSE gap refers to HSE-relaxed band gap, SOC gap refers to HSE-relaxed with SOC band gap, $\Delta$(HSE-PBE) is the difference between PBE and HSE-relaxed band gap, and $\Delta$(SOC-HSE) is the difference between the HSE-relaxed band gap with and without SOC.}
\end{figure}

Finally, we examine the relationships between HaP composition and various types of band gaps discussed above, by calculating Pearson coefficients of linear correlation. Fig \ref{fig:pearson_HSE} shows the correlations for five types of properties, namely E$_{gap}$$^{PBE}$ (PBE Gap), E$_{gap}$$^{HSE-rel}$ (HSE Gap), E$_{gap}$$^{HSE-rel-SOC}$ (SOC Gap), E$_{gap}$$^{HSE-rel}$ -- E$_{gap}$$^{PBE}$ ($\Delta$(HSE-PBE)), and E$_{gap}$$^{HSE-rel-SOC}$ -- E$_{gap}$$^{HSE-rel}$ ($\Delta$(SOC-HSE)). For the first three quantities, we find virtually identical behavior, and it can be concluded that various A/B/X-site species have the same increasing or decreasing influence on any PBE or HSE E$_{gap}$. Correlations with the (HSE -- PBE) E$_{gap}$ difference values show that certain constituents such as FA, Cs, Sn, or I could have marginal influence, but the differences are largely uniform across the dataset. The effect of SOC is very evident in the correlation analysis for E$_{gap}$$^{HSE-rel-SOC}$ -- E$_{gap}$$^{HSE-rel}$. While A-site species have no influence here, Pb has by far the highest negative correlation, Ge has a slightly positive correlation, and I has a slightly negative correlation. We conclude that inclusion of SOC is of utmost importance for Pb-based HaPs, and E$_{gap}$ values will be significantly lower (and more accurate) when using SOC. Figs. S7, S8, and S9 present the complete correlation analysis for all properties computed from the three types of HSE functionals and the 50-dimensional descriptors introduced earlier, showing very similar trends as compared to the PBE dataset.

\begin{figure*}[h]
\centering
\includegraphics[width=1.0\linewidth]{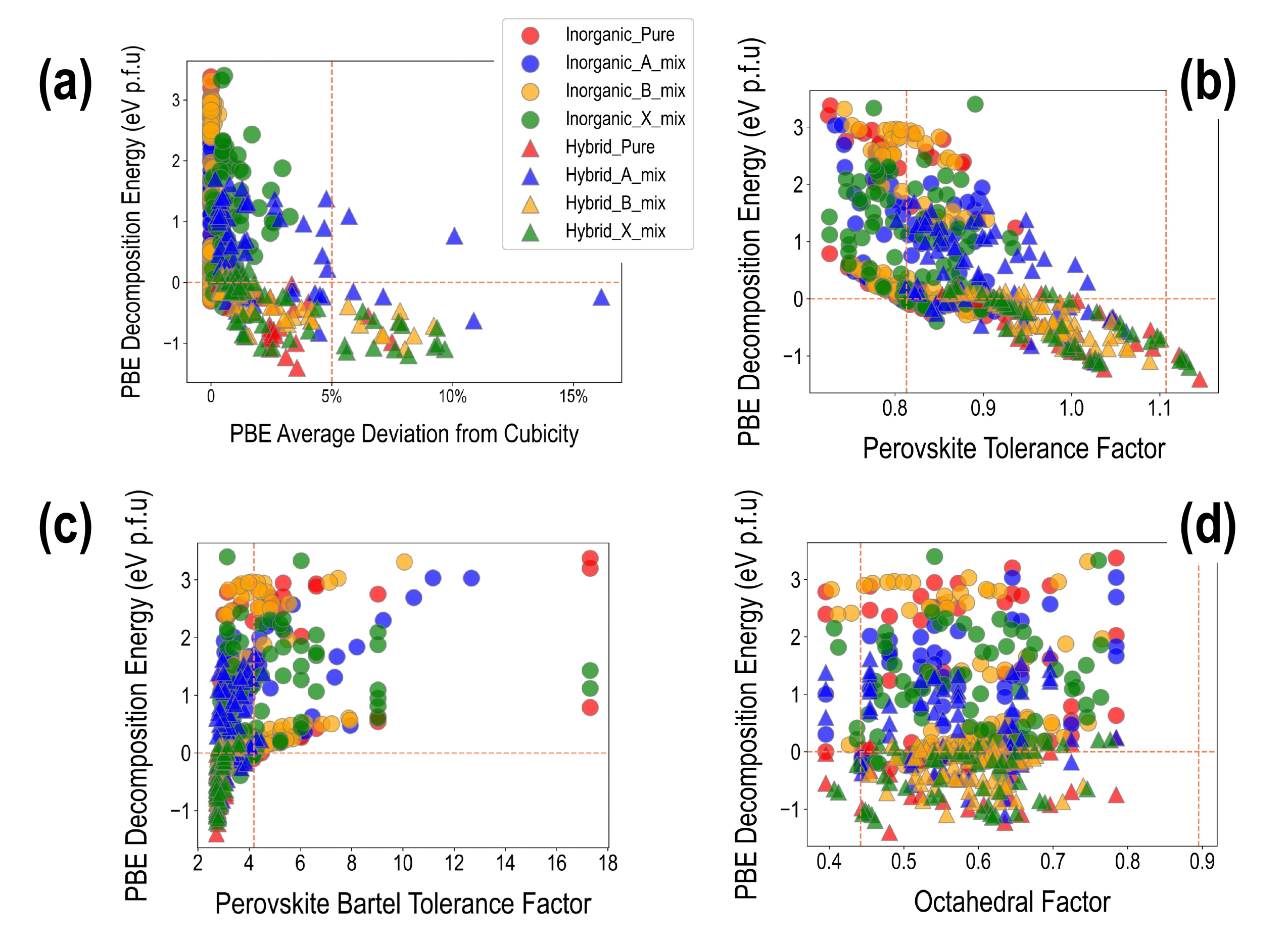}
\caption{\label{fig:cuts} The PBE computed decomposition energy plotted against (a) average deviation from cubicity, (b) Goldschmidt tolerance factor, (c) Bartel tolerance factor, and (d) octahedral factor.}
\end{figure*}

\subsection{Deviation from Cubicity}

As alluded to earlier, some of the PBE and/or HSE geometry optimization calculations, especially when multiple organic cations are present in the HaP supercell, could lead to significant distortions of the cubic perovskite structure. Beyond the use of a perfectly cubic supercell as the starting geometry, the cubic shape is not enforced in the computations, but a vast majority of the structures in the dataset are cubic or pseudo-cubic. Here, we investigate how far any structure deviates from an acceptable pseudo-cubic shape, and use this information to subsequently screen out severely deformed, visibly non-cubic, or unphysical perovskite phases, even if the energy may be low. We define a metric known as Deviation from Cubicity (DC), estimated by how different $\bar{b}$ and $\bar{c}$ lattice constant values are compared to lattice constant a, as shown in equation \ref{eq:cubicity_l}. Similarly, an angular deviation is calculated by measuring how different angles $\alpha$, $\beta$, and $\gamma$ are from 90 degrees, as shown in equation \ref{eq:cubicity_aplha}. DC values greater than 10\% for lattice constant and greater than 5\% are considered to be too non-cubic and excluded during the screening process, which will be explained in a later section.

\begin{equation}\label{eq:cubicity_l}
DC_b=\frac{|b-a|}{a}
\end{equation}

\begin{equation}\label{eq:cubicity_aplha}
DC_\alpha=\frac{|\alpha-90^{\circ}|}{90^{\circ}}
\end{equation}

\begin{equation}\label{eq:cubicity_avg}
DC_{avg}=\frac{DC_b+DC_c+DC_\alpha+DC_\beta+DC_\gamma}{5}
\end{equation}

Fig \ref{fig:cuts}(a) shows $\Delta$H$^{PBE}$ plotted against the average deviation from cubicity (DC$_{avg}$), calculated using equation \ref{eq:cubicity_avg}. Figs. S10 and S11 show individual plots of $\Delta$H$^{PBE}$ against DC corresponding to $\bar{b}$, $\bar{c}$, $\alpha$, $\beta$, and $\gamma$. It can be seen that $\sim$ 90\% of the compounds show DC$_{avg}$ of < 2\%, reinforncing confidence in the cubic/pseudo-cubic nature of a majority of the dataset. Around 20 compounds show DC$_{avg}$ of > 5\%, and all of them are hybrid HaPs with A-site, B-site, or X-site mixing. The non-symmetricity introduced in the supercell when large organic molecules are mixed with other organic or inorganic cations, and when complex mixing is performed at the B or X sites in the presence of large organic cations, leads to elongation, contraction, or twist along one or more directions. A consequence of the high-throughput nature of our computational work is the inability to visually examine every structure and its likeliness to a perovskite cubic phase: the current analysis helps reveal some unfavorable deviations in certain compounds, which will be used as one of the factors while determining suitable compositions in terms of perovskite formability, stability, and optoelectronic properties.

\begin{figure*}[t]
\centering
\includegraphics[width=1.0\linewidth]{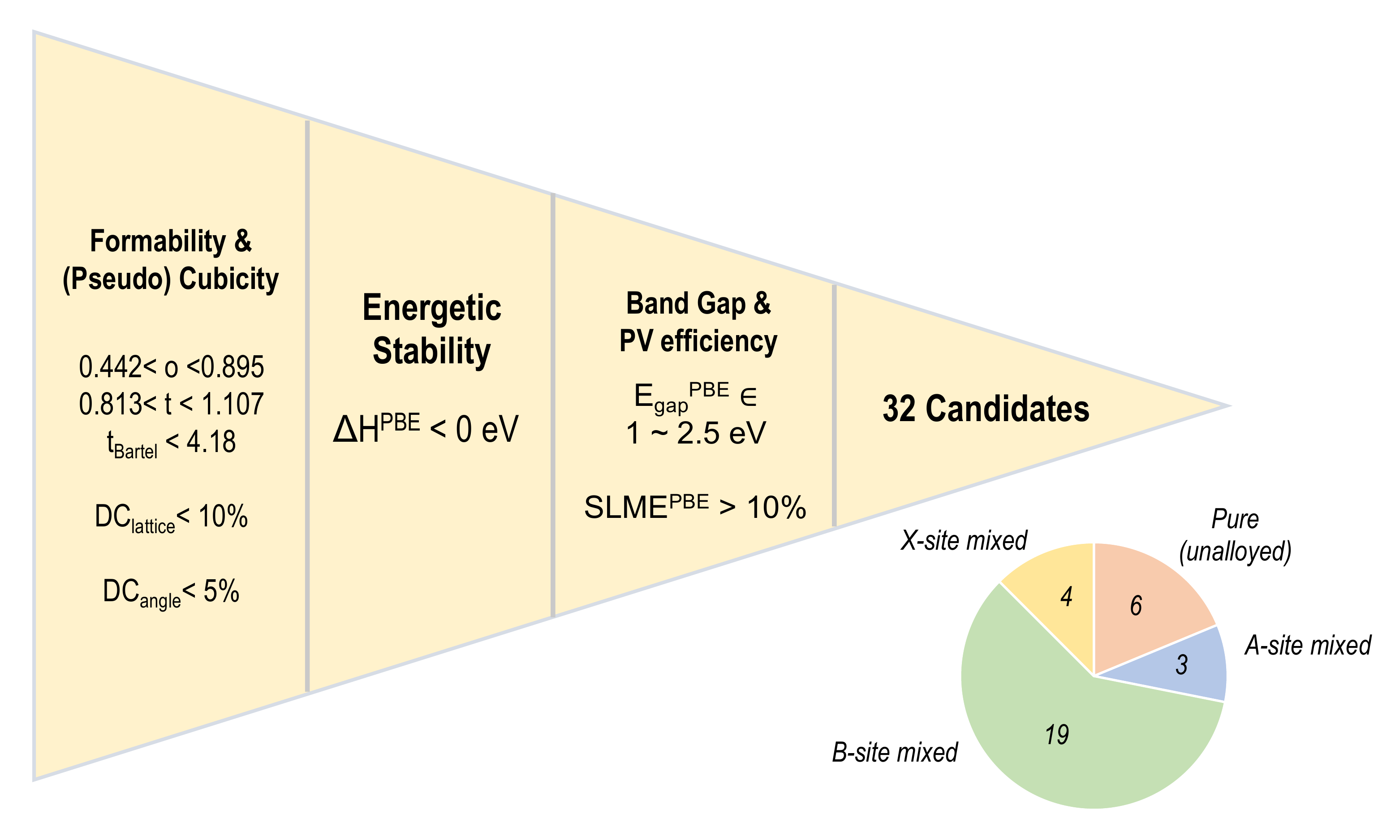}
\caption{\label{fig:PBE_funnel} Screening methodology applied on the PBE dataset, yielding 32 candidates that satisfy all perovskite formability and cubicity conditions, show negative decomposition energy, and PV-appropriate band gap and SLME. The pie chart shows the distribution of various alloy types in the screened list of compounds.}
\end{figure*}

\subsection{Comparing Perovskite Formability Factors with Decomposition Energy}

The formability of an ABX$_3$ perovskite is typically predicted using the Goldschmidt tolerance and octahedral factors, which depend on the ionic radii of A, B, and X-site species. In recent years, there have been newer factors devised through analysis of large quantities of experimental and computational perovskite data, often using machine learning techniques \cite{MRS_Bulletin_JY_Arun}; one such factor was suggested by Bartel et al. \cite{bartel-2019-new-toler} Here, we utilize three factors, namely the traditional tolerance factor (t), octahedral factor (o), and the Bartel tolerance factor ($t_B$), defined using equations \ref{eq:factor_o}, \ref{eq:factor_t} and \ref{eq:factor_b} respectively, to quantify the formability of all perovskites in our dataset and compare these values with DFT computed $\Delta$H. For compounds with mixing, weighted averages of A-site ($r_A$), B-site ($r_B$), and X-site ($r_X$) radii are considered.

Octahedral factor:
\begin{equation}\label{eq:factor_o}
o=\frac{r_B}{r_X}
\end{equation}

Tolerance factor: 
\begin{equation}\label{eq:factor_t}
t=\frac{r_A+r_X}{\sqrt{2}(r_B+r_X)}
\end{equation}

Bartel\cite{bartel-2019-new-toler} tolerance factor: 
\begin{equation}\label{eq:factor_b}
t_{Bartel}=\frac{r_X}{r_B}-[1-\frac{\frac{r_A}{r_B}}{ln(\frac{r_A}{r_B})}]
\end{equation}

\begin{figure*}[t]
\centering
\includegraphics[width=1.0\linewidth]{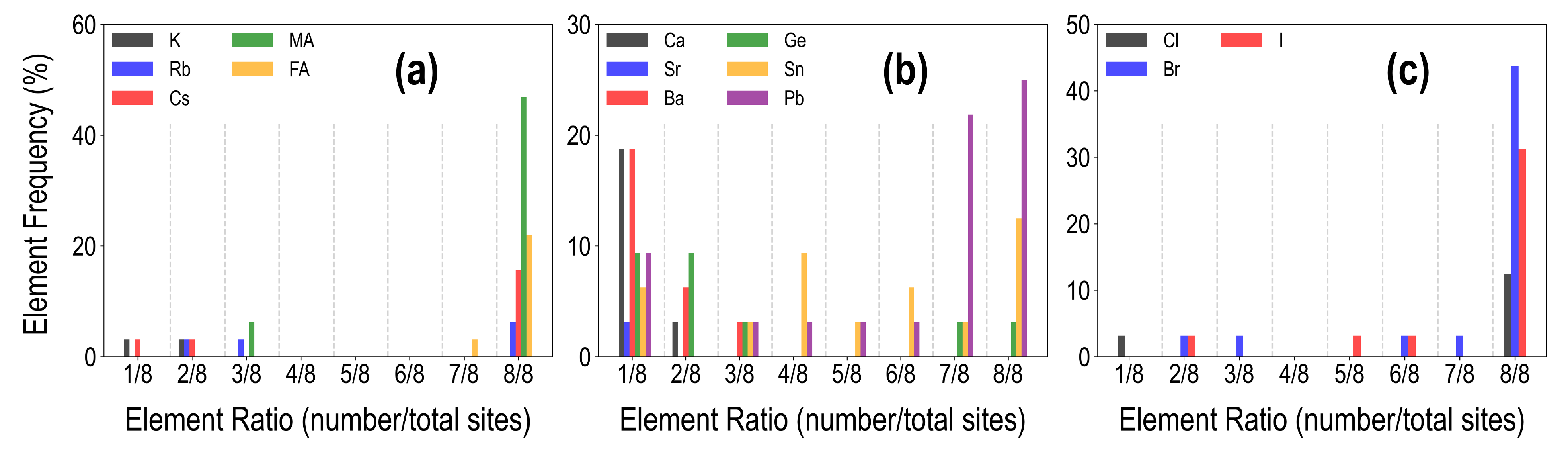}
\caption{\label{fig:PBE_freq_screen} Distribution of mixing fractions of various species at the A (a), B (b), and X (c) sites across the list of 32 promising compounds selected from the PBE dataset.}
\end{figure*}

The suggested ranges for perovskite formability are o $\in$ (0.442 - 0.895), t $\in$ (0.813 - 1.107), and $t_B$ < 4.18. Figs. \ref{fig:cuts}(b), \ref{fig:cuts}(c), and \ref{fig:cuts}(d) respectively show t, $t_B$, and o values for the entire dataset plotted against $\Delta$H$^{PBE}$. It can be seen that there is a roughly inverse relationship between t and $\Delta$H$^{PBE}$, which is to be expected as larger values of t mean more favorable perovskite formability and should thus also correspond to negative decomposition energies. A small number of compounds lie in the t > 1.1 range and also show negative $\Delta$H$^{PBE}$; upon closer inspection, we find that these are hybrid perovskites with significantly distorted structures showing large deviation from cubicity. One example of such a compound is FAGeBr$_{2.25}$Cl$_{0.75}$, which has t = 1.13 and $\Delta$H$^{PBE}$ = -1.06 eV, but a highly distorted PBE optimized structure with DC$_b$ = 16.4\%. Fig. \ref{fig:cuts}(c) shows that essentially all HaP compositions with negative $\Delta$H$^{PBE}$ fall within the desirable $t_B$ < 4.18 region. Similarly, Fig. \ref{fig:cuts}(d) shows that nearly all compounds with negative $\Delta$H$^{PBE}$ lie in the desirable range of o values, barring a very small number of distorted structures. It should be noted from Figs. \ref{fig:cuts}(b), (c), and (d) that hundreds of compositions that satisfy the formability factor conditions show positive (often very large positive) $\Delta$H$^{PBE}$ values, which means they will easily decompose to other halide phases. Our observations point to the idea that such factors may be necessary but not sufficient conditions for perovskite formability and stability.

\subsection{High-Throughput Screening of Compositions with Favorable Properties}

So far, we have visualized and analyzed an HT-DFT dataset of HaP alloys from PBE and multiple HSE functionals. In performing an initial screening of candidates with promise for single-junction solar absorption, we must consider $\Delta$H (a necessary but not complete description of perovskite stability), E$_{gap}$, and SLME; in addition, established perovskite formability factors as well as deviation from cubicity should be considered. We note here that the DFT dataset covers as wide a compositional spectrum of HaPs as possible, within the 14-dimensional chemical space. There are, of course, innumerable compositions that could be generated which are intermediate to those currently being studied, such as by mixing in fractions other than n/8 (where n is a positive integer), which essentially implies simulations in larger supercells, and which may lead to even more desirable combinations of properties. We tackle this issue in future works by building upon our current dataset and applying state-of-the-art ML algorithms for prediction and inverse design. For the moment, we use the criteria/properties described in the previous section to screen for promising materials within the DFT datasets.

We apply the following screening criteria on the PBE dataset:

\begin{enumerate}
    \item Formability: o $\in$ {0.442 - 0.895}, t $\in$ {0.813 - 1.107}, and $t_B$ < 4.18
    \item (Pseudo) cubicity: DC$_{b}$ < 10\%, DC$_{c}$ < 10\%, DC$_{\alpha}$ < 5\%, DC$_{\beta}$ < 5\%, and DC$_{\gamma}$ < 5\%, 
    \item Thermodynamic stability: $\Delta$H$^{PBE}$ < 0 eV. Negative $\Delta$H is a necessary but not complete metric for preventing ABX$_3$ decomposition to phases AX and BX$_2$; decomposition could happen to other phases, and the effects of kinetics, ion segregation, defects, etc. are ignored in this work. 
    \item Band gap: E$_{gap}$$^{PBE}$ $\in$ {1 eV - 2.5 eV}. PV-suitable band gaps lie close to 1.5 eV. We use a wide range here to account for the various inadequacies of the PBE band gap description: it will underestimate gaps of inorganic compounds but either accidentally be accurate or slightly overestimate gaps of hybrid HaPs: this effect has been studied in prior works. \cite{mannodi-kanakkithodi-2019-compr-comput,mannodi-kanakkithodi-2022-data-driven}
    \item PV efficiency: SLME$^{PBE}$ > 0.10. This criterion goes hand-in-hand with the band gap requirement, as it can be seen from Fig \ref{fig:pairplots}(b) that the highest SLME values correspond roughly to the band gap range described above.
\end{enumerate}

\begin{figure}[t]
\centering
\includegraphics[width=1.0\linewidth]{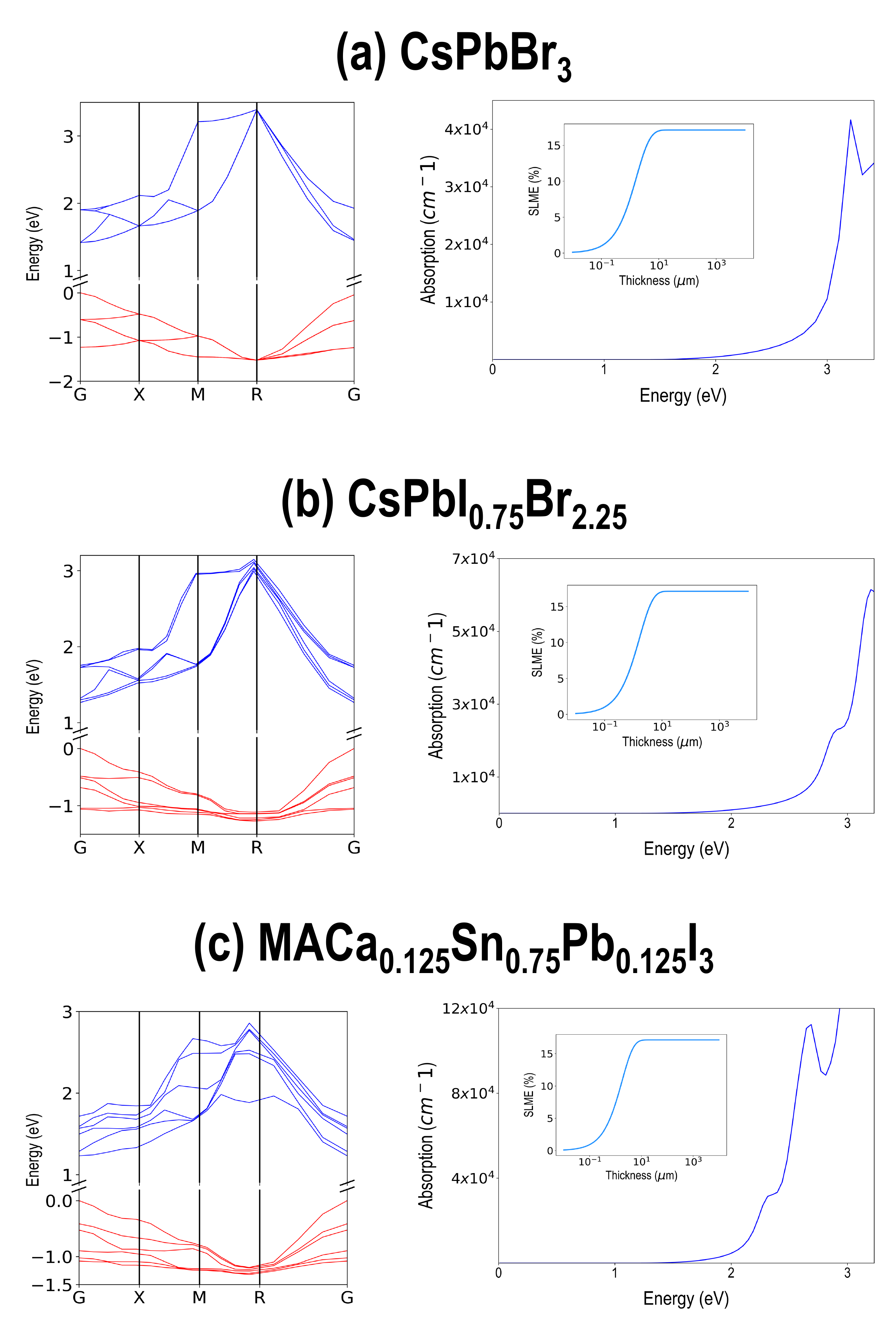}
\caption{\label{fig:top_3} HSE-PBE-SOC calculated electronic band structures, optical absorption spectra, and SLME vs sample thickness (inset) plots for three promising compounds, namely (a) CsPbBr$_3$, (b) CsPbI$_{0.75}$Br$_{2.25}$, and (c) MACa$_{0.125}$Sn$_{0.75}$Pb$_{0.125}$I$_3$.}
\end{figure}

Fig. \ref{fig:PBE_funnel} shows our five-fold screening process, based on which we obtain 32 candidates (out of 495) that fulfill each requirement. Also shown is a pie chart with the distribution of various types of mixing in the screened list of compounds. It can be seen that a majority of the screened compounds--19 in total--are B-site mixed, and there are only 6 un-alloyed compositions. Fig. \ref{fig:PBE_freq_screen} further shows the relative distributions of various A-site, B-site, and X-site species. We find that MA followed by FA are by far the most common A-site cations, often occupying all of the A-site (8/8 mixing fraction), followed by Cs, and Rb. There are no pure K-based compounds; K, as well as Rb and Cs, occur in small fractions in some of the compounds. Pb and Sn appear in an overwhelming majority of the compounds, with Ge, Ca, Sr, and Ba only occurring in smaller fractions of 3/8 or less. This is consistent with the observation that Pb and Sn, and sometimes Ge, are most beneficial for ideal optoelectronic properties, whereas Ca/Sr/Ba should occur in minor fractions to keep the band gap small. Pb has a high preference for 7/8 and 8/8 occupation, hinting at the difficulty in developing Pb-free perovskites with ideal properties. At the X-site, Br and I without any mixing are most common, and in the 4 compounds with X-site mixing, I, Br, and Cl are found in various fractions.

We performed a similar screening procedure using the HSE-PBE-SOC dataset, which was found to compare best with experiments for band gap. Applying the very same criteria as shown in Fig. \ref{fig:PBE_funnel} leads to a list of 14 stable and formable compounds with desirable band gaps and SLME, out of which 4 are pure unalloyed compounds, 1 is (purely inorganic) A-site mixed, 8 are B-site mixed, and 1 is X-site mixed. Distributions of the types of mixing fractions of various species in this screened list of compounds is shown in Fig. S12. Although the HSE-PBE-SOC screened list is much smaller than the PBE screening due to a smaller overall dataset, some similar trends are found in both screening precedures. B-site mixing is most prevalent, as is high fractions of Pb, and sometimes Sn and Ge, at the B-site. Ca/Sr/Ba prefer mixing in small fractions. Most compounds are MA-based and nearly all of them contain Br or I. 

The entire screened lists of compounds from PBE and HSE-PBE-SOC are presented in Tables SII and SIII respectively, along with their chemical formula and (PBE or HSE-PBE-SOC) computed $\Delta$H, E$_{gap}$, and SLME at 5$\mu$m sample thickness. Interestingly, all the compounds in the HSE-PBE-SOC list appear in the PBE list as well. Three of the best performing compounds are selected and their HSE-PBE-SOC computed electronic band structures, optical absorption spectra, and SLME vs sample thickness plots are pictured in Fig. \ref{fig:top_3}. These compounds, namely CsPbBr$_3$, CsPbI$_{0.75}$Br$_{2.25}$, and MACa$_{0.125}$Sn$_{0.75}$Pb$_{0.125}$I$_3$, show direct band gaps around 1.5 eV and SLME > 15\% in their cubic or pseudo-cubic phases.

\section{Perspective and Future Work}

What we reported in this work is one of the largest DFT datasets to date of pseudo-cubic HaP alloys containing some of the most commonly used cation and anion species. This data enabled us to understand the dependence of stability and optoelectronic properties on perovskite composition, specifically the type of mixing. However, this work is but the first step in a very long process that will involve extensions to non-cubic phases, other properties of interest, more improved levels of theory, alternative cation and anion choices, and other perovskite forms such as double perovskites and 2D perovskites, ultimately leading to more universal prediction, screening, and design.

The immediate next extension is towards non-cubic perovskite phases. For instance, CsPbBr$_3$ may prefer the orthorhombic phase, while MAPbI$_3$ and MA(Pb-Sn)I$_3$ may assume tetragonal phase, and this work considers all such compounds only in a cubic or pseudo-cubic rendition. In previous work \cite{mannodi-kanakkithodi-2022-data-driven}, it was shown that for the same composition, unalloyed or with mixing, changing the phase could modify the band gap by 0.5 eV or more in many cases. Non-cubic phases are either the most stable, or metastable/competing phases for most of the compositions studied in this work. Currently, we have high-throughput computations ongoing for tetragonal, orthorhombic, and hexagonal phases of several mixed HaPs; the perovskite phase itself can be added as an input to the compositional and elemental descriptors to obtain new correlations. In addition, computations are being performed for further tailoring of properties by accessing polymorphs within each phase, e.g., via octahedral distortion and rotation \cite{oct}, or via re-optimization of the same composition in larger supercells with slight distortions \cite{Dalpian}. 

We further anticipate significant improvements in DFT predictions of various properties. Our attempt to utilize a few different functionals to benchmark properties against experiments was hindered by a number of factors discussed in the manuscript, including the perovskite phase and lack of additional corrections. Our ongoing computations involve testing the influence of the PBEsol \cite{PBEsol} and PBE-D3 \cite{PBE-D3} functionals, combined with static HSE06 or GW computations with SOC \cite{perovs_GW}, for better optical and electronic properties. Furthermore, the inclusion of new types of elemental or molecular species--such as transition metals (Cd, Zn, Ni, etc.) at the B-site--would necessitate the use of specific levels of theory, such as GGA+U \cite{gga_u}. Consideration of other important properties, such as defect formation energies and carrier mobilities, would involve testing and deploying multiple functionals as well. 

It should be noted again that there might not be one best functional that works for the entire chemical space when considering organic vs inorganic A-site cations, and Pb/Sn vs other B-site cations. A likely solution is the use of an ensemble of functionals as well as experimental estimates (which might need to be averaged as well, given the range of values generally reported by different experimental researchers for the same materials) for hundreds of HaP compositions, and training of multi-fidelity ML models \cite{mfml}. Large quantities of low-fidelity data combined with more modest amounts of high-fidelity data can lead to highly accurate predictions of experiment-level property estimates.

In general, ML has a massive role to play here, as has been demonstrated for HaPs in multiple prior works \cite{mannodi-kanakkithodi-2022-data-driven,MRS_Bulletin_JY_Arun}. Concurrent manuscripts are planned to report rigorously optimized predictive models for multiple properties and fidelities, based on the datasets and descriptors discussed in this work. Such models can easily be extended to new choices for A/B/X ions such as transition metals \cite{mannodi-kanakkithodi-2022-data-driven}, as well as other phases, by addition of new dimensions to the descriptors. The inclusion of more general crystalline structure representations as inputs for ML, such as using crystal graphs and graph neural networks \cite{CGCNN,MEGNET,ALIGNN}, would be essential for treating same compositions and structures with a variety of distortions or lattice strains. Finally, inverse design techniques, such as using genetic algorithm \cite{GA} or generative neural networks \cite{GAN}, could be applied upon the DFT-ML surrogate models to drive the efficient discovery of new HaP compositions/structures with multiple desired properties. The dataset and analysis presented in this work serves as a springboard for efforts that are currently underway, to ultimately accelerate the prediction and design of novel perovskites for optoelectronics, and to extend such approaches to other material classes and applications.

\section{Conclusions}

In this work, we present a high-throughput DFT dataset of pseudo-cubic ABX$_3$ halide perovskite alloys, with mixing of multiple ions permitted at the A, B, or X sites, using the GGA-PBE functional and three types of hybrid HSE06 approaches. This dataset contains 495 unique compositions with PBE computed decomposition energies, band gaps, and spectroscopic maximum limited efficiencies (SLME) from the optical absorption spectra, and the same properties for 299 compounds from full HSE relaxation, 282 compounds from HSE relaxation with spin-orbit coupling (SOC), and 244 compounds from static HSE computations on PBE relaxed structures with SOC. Pearson correlation analysis reveals the extent of positive or negative correlation of the amount of any A/B/X species as well as their well known elemental/molecular properties on the computed stability and optoelectronic properties, reproducing known trends and unraveling interesting new relationships. Screening is performed for materials resistant to decomposition, with photovoltaic-suitable band gaps and high SLME, as well as including other perovskite formability factors such as Goldschmidt tolerance and octahedral factors and the deviation of perovskite structure from cubicity, to obtain 32 promising compounds from PBE and 14 from HSE-PBE-SOC. This works forms the basis for predictive machine learning models which will accelerate the design of novel perovskites with attractive properties.

\section*{Conflicts of Interest}
There are no conflicts to declare.

\section*{Data Availability}
All raw DFT data, including input and output files, can be found on Materials Data Facility \cite{MDF}. Tabulated data in the form of chemical formulas and properties computed from various functionals is included as .xlsx files in the supporting documents. The tabulated data and scripts used to analyze DFT calculated properties can be found at https://github.com/yjq829/perovskite\_dataset.git.

\section*{Acknowledgements}
Extensive discussions with and scientific feedback from Prof. David Fenning (UC San Diego), Dr. Rishi Kumar (Berkeley lab), and Dr. Maria Chan (Argonne National Lab) are acknowledged. This work was performed at Purdue University, under startup account F.10023800.05.002 from the Materials Engineering department. This research used resources of the National Energy Research Scientific Computing Center, the Laboratory Computing Resource Center at Argonne National Laboratory, and the RCAC clusters at Purdue.

\bibliographystyle{rsc}
\bibliography{main}

\clearpage
\newpage
\pagenumbering{gobble}
\thispagestyle{empty} 

\onecolumn

\setcounter{figure}{0}   
\setcounter{table}{0} 
\renewcommand{\thetable}{S\Roman{table}} 
\renewcommand\thefigure{S\arabic{figure}}

\begin{center}
\vspace{0.5cm}
\Large
\textbf{Supplemental material to "A High-Throughput Computational Dataset of Halide Perovskite Alloys"\\}
\vspace{0.5cm}
\large
Jiaqi Yang,\textsuperscript{1)}, Panayotis Manganaris,\textsuperscript{1)}, and Arun Mannodi-Kanakkithodi \textsuperscript{1, a)}\\
\vspace{0.3cm}

\normalsize
\textsuperscript{1}\textit{School of Materials Engineering, Purdue University, West Lafayette, Indiana 47907, USA}\\
\end{center}

\footnote{
\textsuperscript{a}amannodi@purdue.edu\hspace{0.3cm}}

\vspace{1cm}

\thispagestyle{empty}

\begin{figure*}[h]
\centering
\includegraphics[width=.9\linewidth]{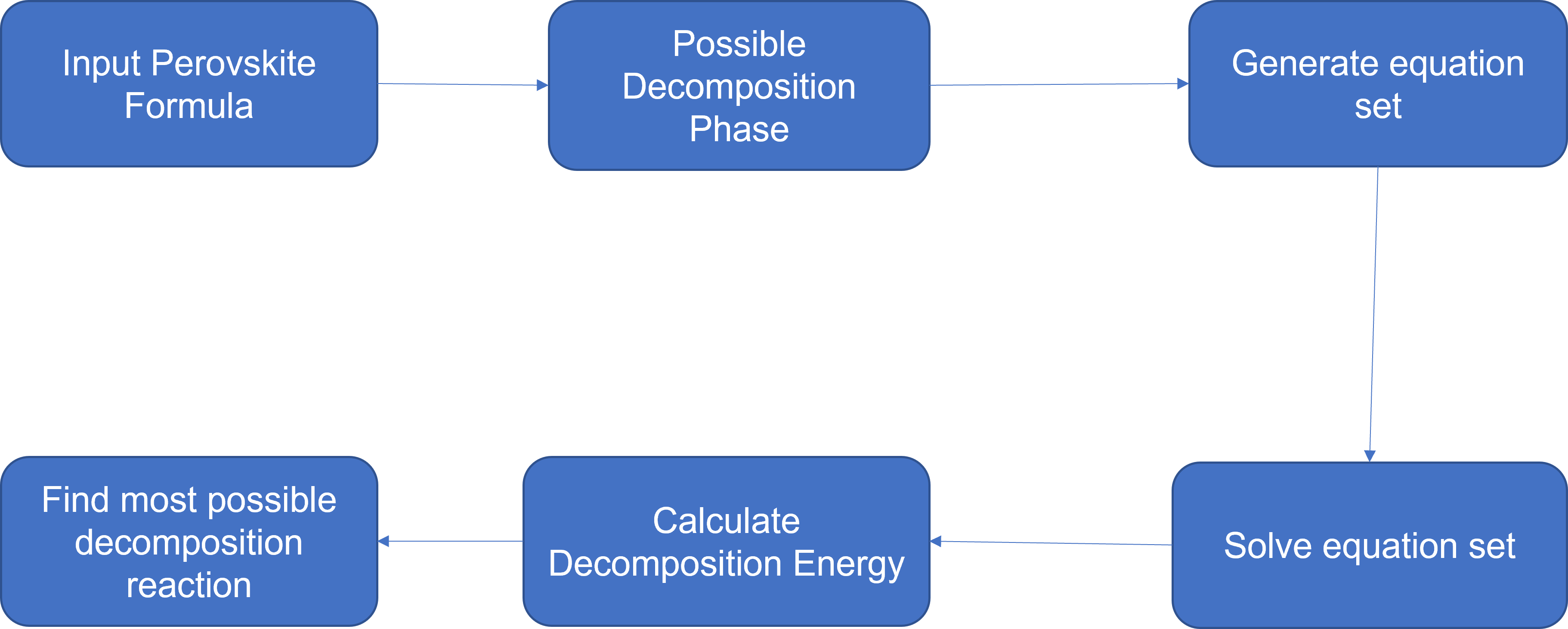}
\caption{\label{fig:decomE_corr_flow} Scheme for determining the perovskite composition energy.}
\end{figure*}

\begin{figure*}[h]
\centering
\includegraphics[width=.9\linewidth]{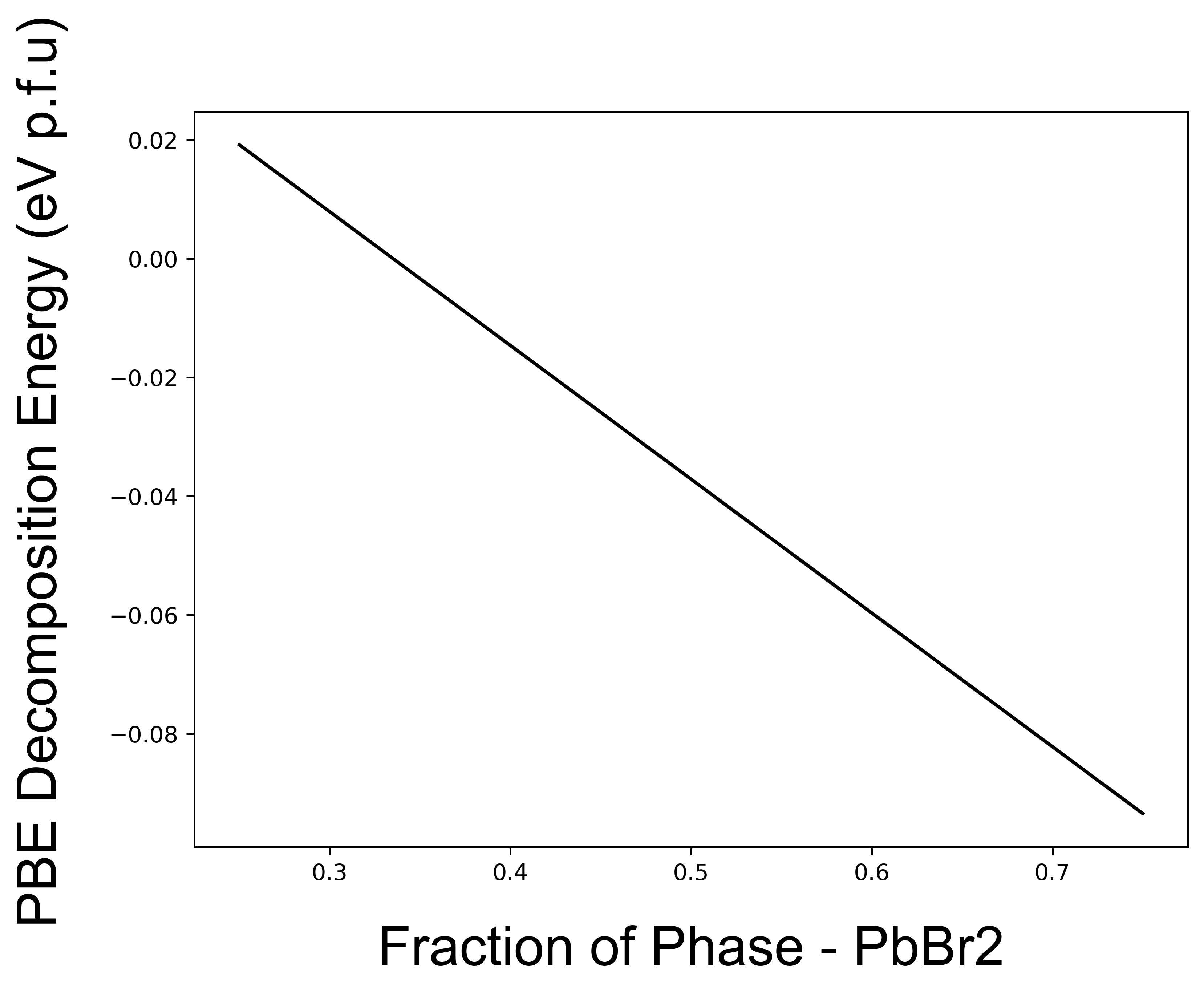}
\caption{\label{fig:decomE_corr_MAPbBrI_example} Decomposition Energy of MAPbBr$_{1.5}$I$_{1.5}$ plotted against the fraction of $PbBr_2$.}
\end{figure*}

\section*{Decomposition Energy Corrections for X-site Mixed Perovskites}

Stability with respect to decomposition is one of the most critical properties of halide perovskites. In this research, we define a ``decomposition energy" to demonstrate the ease or difficulty of a compound to stay as a perovskite (pseudo-cubic) structure and not decompose to other phases. The formulas used are shown in equations \eqref{eq:decoE} and \eqref{eq:decoE_2}. However, for X-site mixed perovskites, the ratio of decomposed phase is not a unique solution. The fraction of each decomposed phase is varied in a large range, which makes the decomposition energy variable. For instance, $MAPbBr_xI_y$ has 4 possible sets of decomposed phases. The decomposition reaction can be shown using equation \eqref{eq:decomE_corr_1}, and the relationship between the fractions $\alpha$ and $\beta$ can be calculated by solving equation set \eqref{eq:decomE_corr_1}.

\begin{equation}\label{eq:decomE_corr_reaction}
\begin{aligned}
MAPbBr_xI_y -> \alpha MABr + (1-\alpha) MAI + \beta PbBr_2 + (1-\beta) PbI_2 \\
\end{aligned}
\end{equation}

\begin{equation}\label{eq:decomE_corr_1}
\begin{aligned}
\alpha + 2\beta = x \\
(1-\alpha)+2(1-\beta)=y \\
\alpha , \beta \in [0,1]\\
\end{aligned}
\end{equation}

To solve this problem, we follow the workflow shown in Fig. S1. To determine which decomposition reaction is most likely, we find the reaction with the highest decomposition energy. Taking $MAPbBr_{1.5}I_{1.5}$ as an example, the PBE decomposition energy is plotted versus the fraction of $PbBr_2$ phase (value of $\beta$) in Fig. S2. All other decomposed phases ($MABr$, $MAI$, $PbI_2$) can be represented using the fraction of $PbBr_2$. The most likely decomposition reaction turns out to be one that will form $MABr$, $PbBr_2$ and $PbI_2$; no MAI will be formed. Using this technique, we apply a correction for all X-site mixed perovskite in our data set. In the future, a code will be developed and released to fit any complex mixed perovskites and find the decomposition reaction with the highest decomposition energy and probability. The results of these corrections also show us a way to increase the stability of a perovskite: by increasing the amount of decomposed phase in the environment, the decomposition reaction will be pushed to left side (perovskite side) and thus stabilize the bulk material.

\begin{figure*}[h]
\centering
\includegraphics[width=1.0\linewidth]{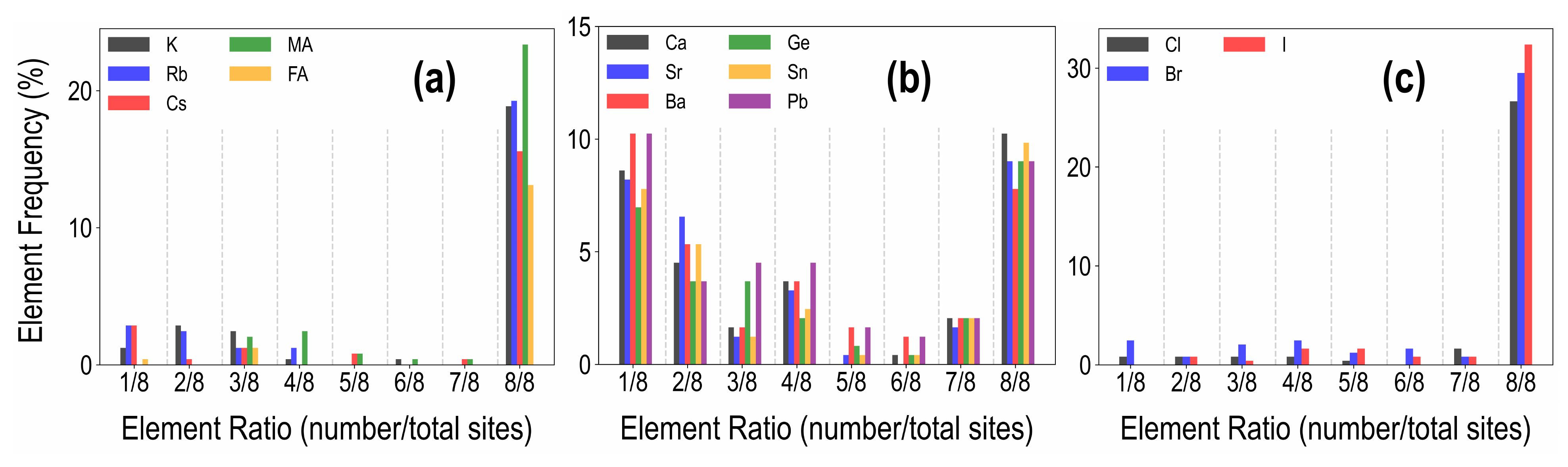}
\caption{\label{fig:HSE-PBE-SOC_freq} Distribution of mixing fractions of various species at the A (a), B (b), and X (c) sites across HSE-PBE-SOC dataset.}
\end{figure*}

\begin{figure*}[h]
\centering
\includegraphics[width=.9\linewidth]{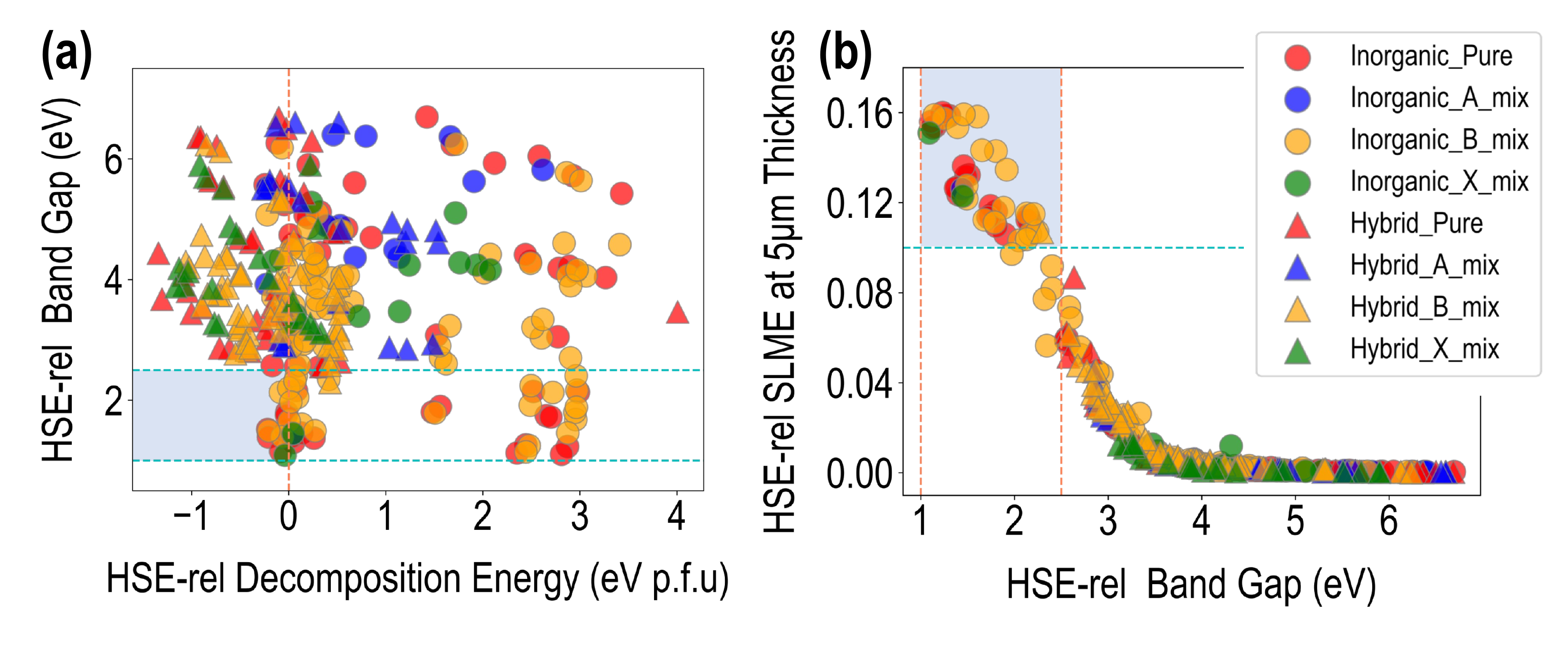}
\caption{\label{fig:HSE_prop} Visualization of the HSE-relaxed dataset: (a) band gap against decomposition energy, and (b) SLME at 5$\mu$m sample thickness against band gap.}
\end{figure*}

\begin{figure*}[h]
\centering
\includegraphics[width=.9\linewidth]{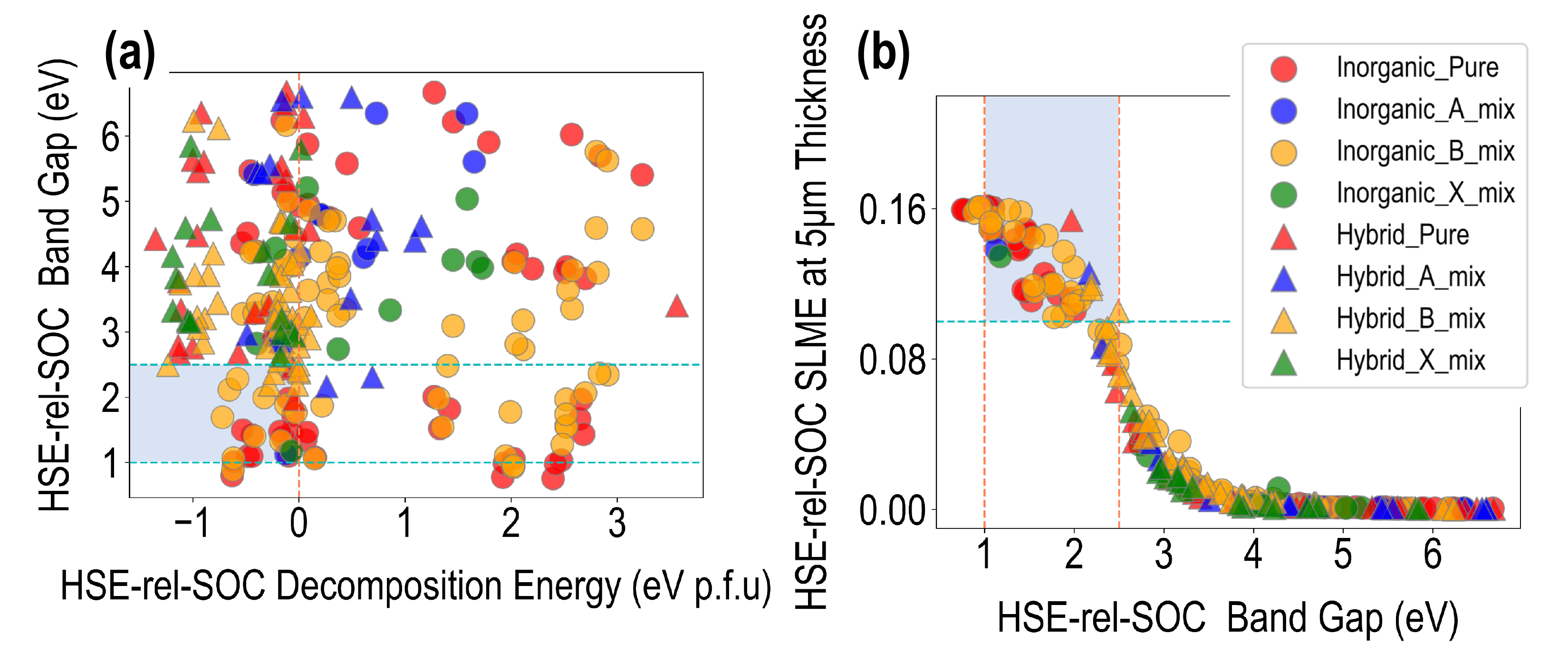}
\caption{\label{fig:HSE_prop} Visualization of the HSE-relaxed with SOC dataset: (a) band gap against decomposition energy, and (b) SLME at 5$\mu$m sample thickness against band gap.}
\end{figure*}

\begin{figure*}[h]
\centering
\includegraphics[width=.9\linewidth]{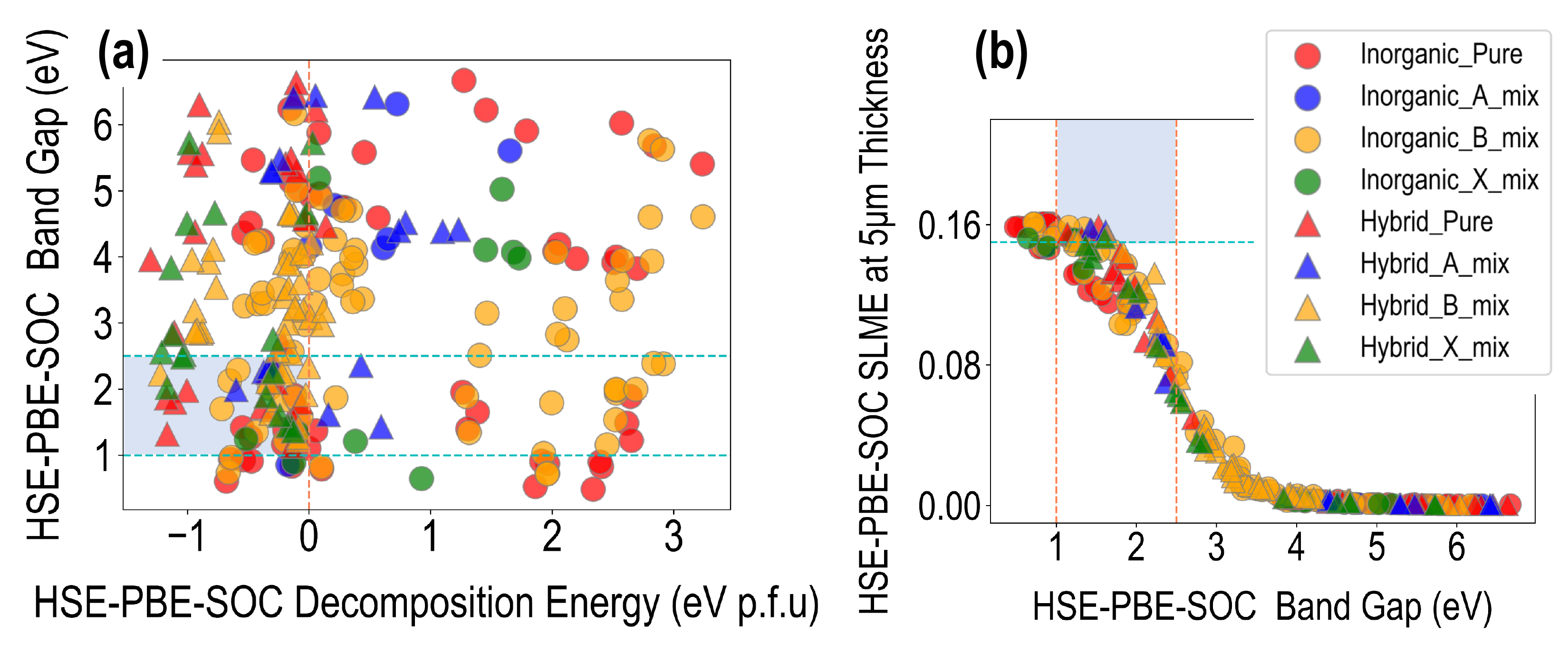}
\caption{\label{fig:HSE_prop} Visualization of the HSE-PBE-SOC dataset: (a) band gap against decomposition energy, and (b) SLME at 5$\mu$m sample thickness against band gap.}
\end{figure*}

\begin{figure*}[h]
\centering
\includegraphics[width=0.9\linewidth]{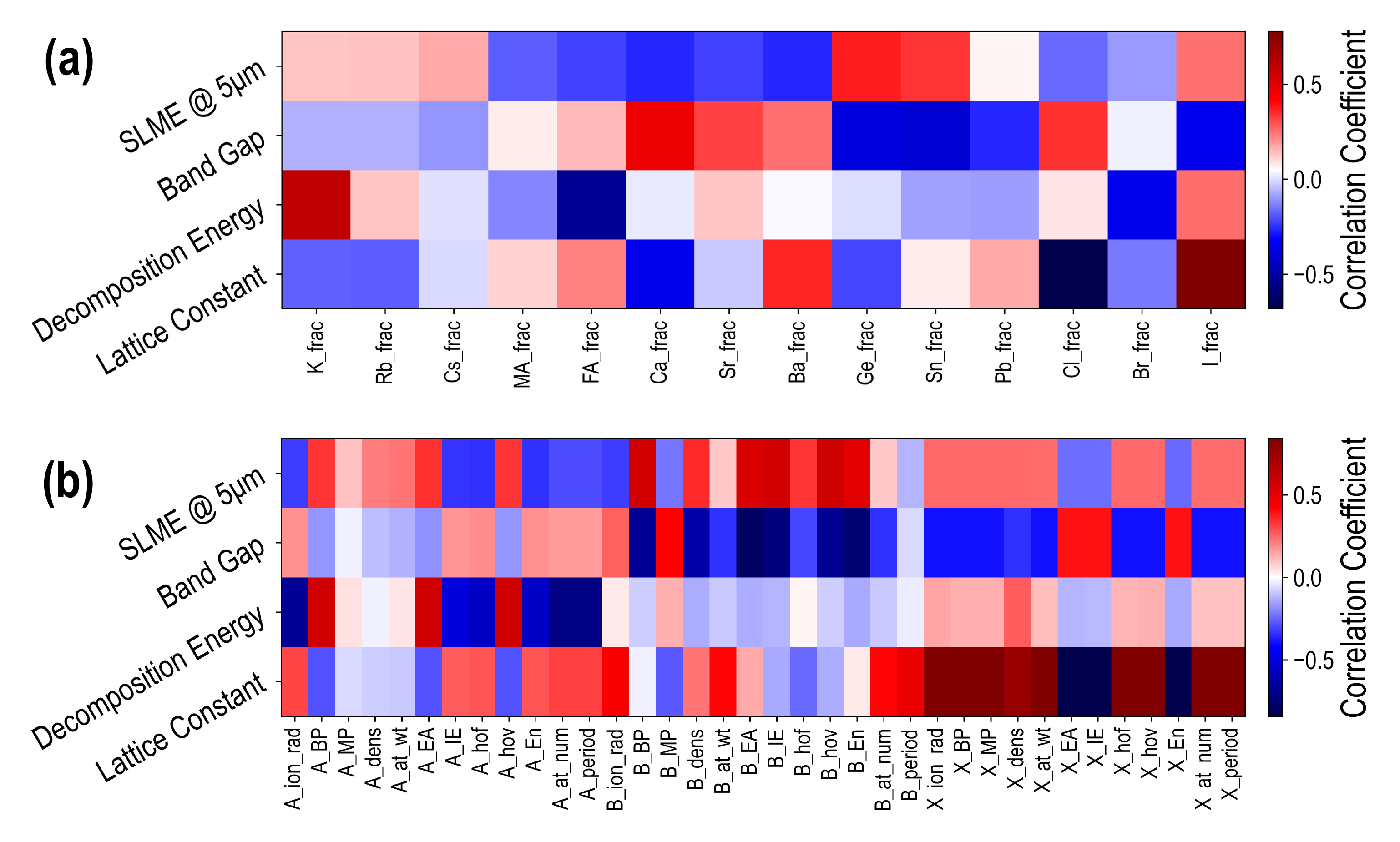}
\caption{\label{fig:pearson_hserel} Pearson coefficients of linear correlation between 4 HSE-relaxed computed properties and (a) 14 compositional descriptors, and (b) 36 elemental property descriptors.}
\end{figure*}

\begin{figure*}[h]
\centering
\includegraphics[width=0.9\linewidth]{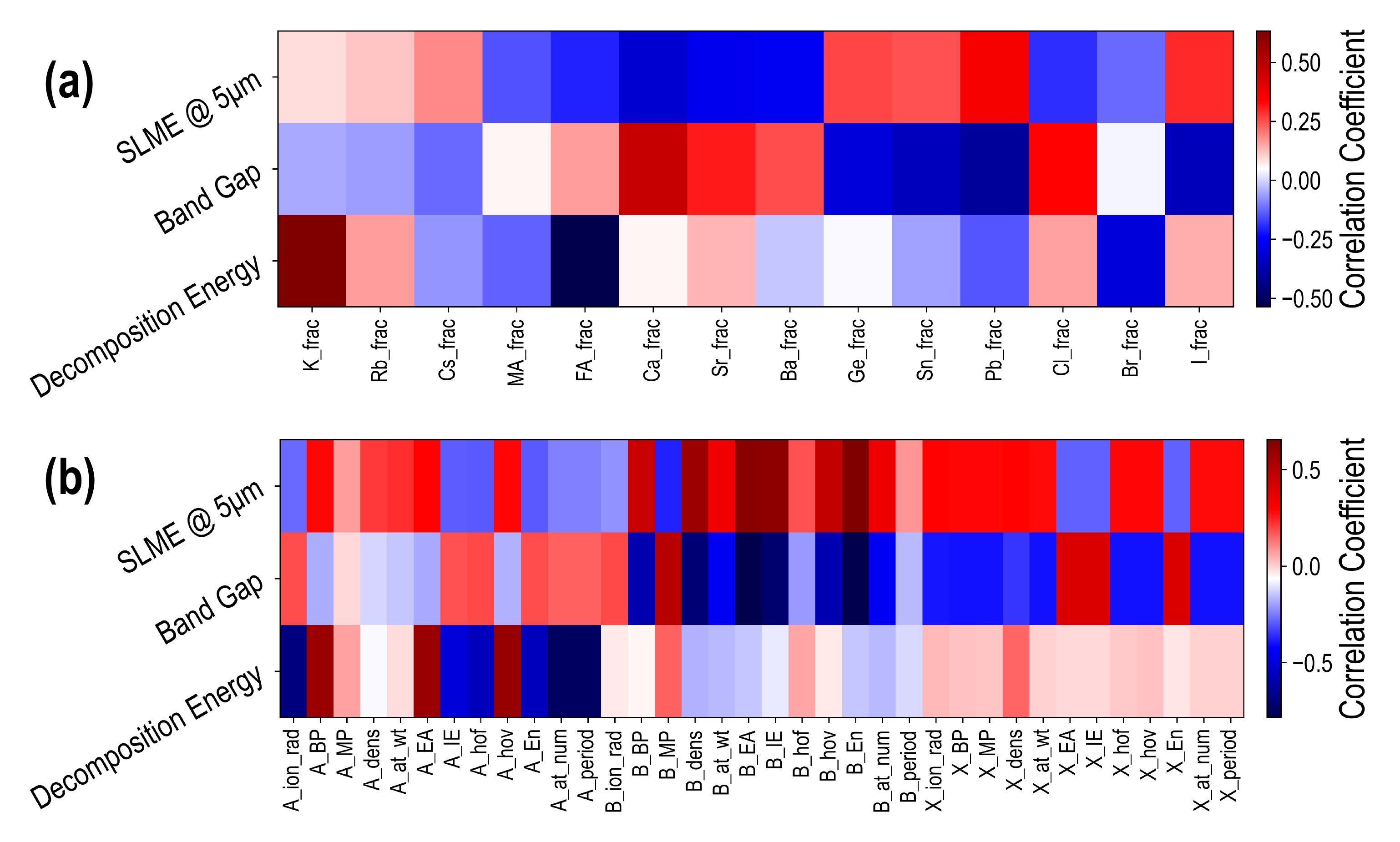}
\caption{\label{fig:pearson_hse_rel_soc} Pearson coefficients of linear correlation between 3 HSE-rel-SOC computed properties and (a) 14 compositional descriptors, and (b) 36 elemental property descriptors.}
\end{figure*}

\begin{figure*}[h]
\centering
\includegraphics[width=0.9\linewidth]{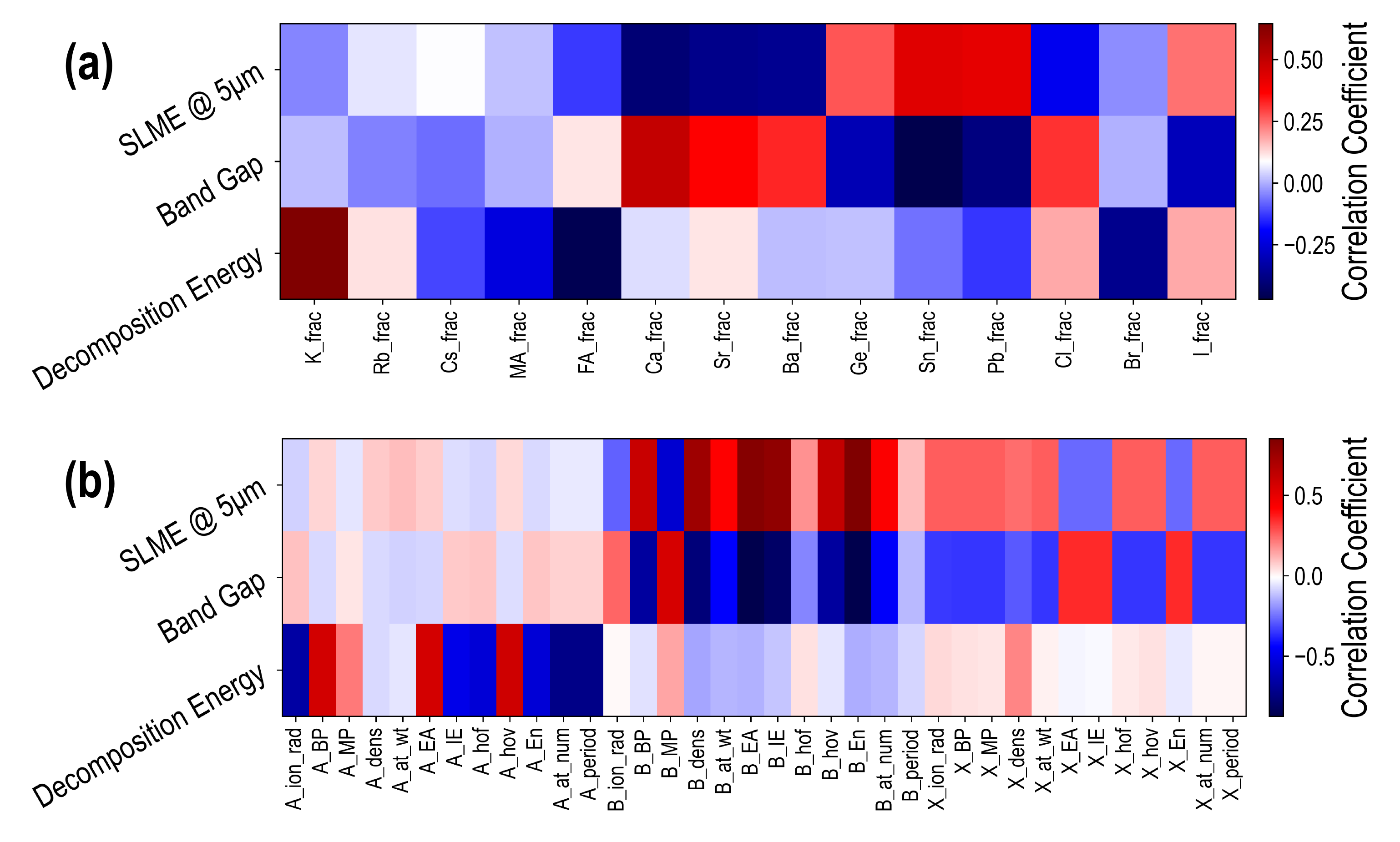}
\caption{\label{fig:pearson_hse_pbe_soc} Pearson coefficients of linear correlation between 3 HSE-PBE-SOC computed properties and (a) 14 compositional descriptors, and (b) 36 elemental property descriptors.}
\end{figure*}

\begin{figure*}[h]
\centering
\includegraphics[width=1.0\linewidth]{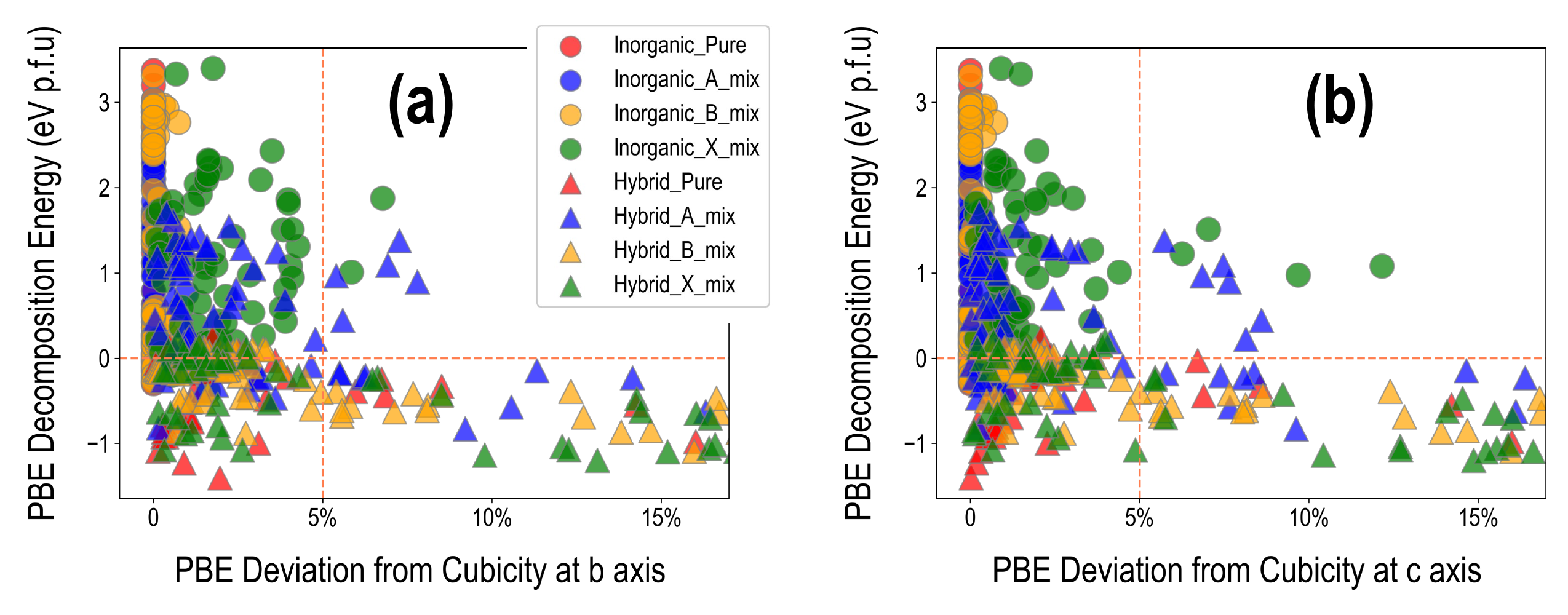}
\caption{\label{fig:cub_b} PBE computed decomposition energy plotted against (a) deviation from cubicity along the b-axis, and (c) deviation from cubicity along the c-axis.}
\end{figure*}

\begin{figure*}[h]
\centering
\includegraphics[width=1.0\linewidth]{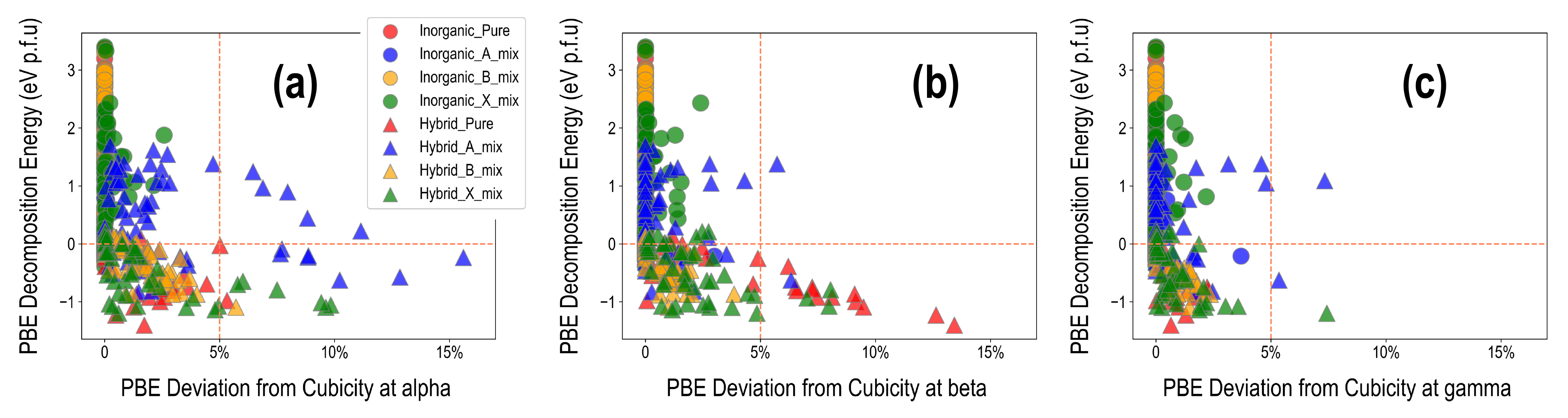}
\caption{\label{fig:cub_alpha} PBE computed decomposition energy plotted against deviation from cubicity in the (a) angle alpha, (b) angle beta, and (c) angle gamma.}
\end{figure*}

\begin{figure*}[h]
\centering
\includegraphics[width=1.0\linewidth]{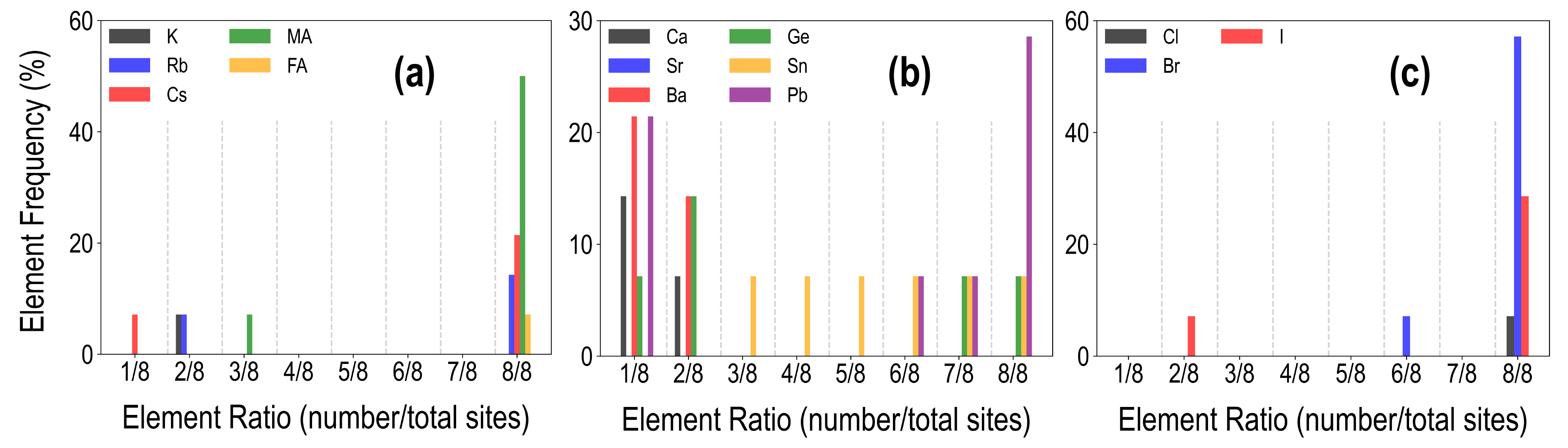}
\caption{\label{fig:HSE-PBE-SOC_freq_screen} Distribution of mixing fractions of various species at the A (a), B (b), and X (c) sites across the list of 14 promising compounds selected from the HSE-PBE-SOC dataset.}
\end{figure*}

\begin{table}[!ht]
    \centering
    \begin{tabular}{|c|c|}
    \hline
        \textbf{Descriptor Label} & \textbf{Descriptor Meaning} \\ \hline
        K\_frac & K fraction over A sites \\ \hline
        Rb\_frac & Rb fraction over A sites \\ \hline
        Cs\_frac & Cs fraction over A sites \\ \hline
        MA\_frac & MA fraction over A sites \\ \hline
        FA\_frac & FA fraction over A sites \\ \hline
        Ca\_frac & Ca fraction over B sites \\ \hline
        Sr\_frac & Sr fraction over B sites \\ \hline
        Ba\_frac & Ba fraction over B sites \\ \hline
        Ge\_frac & Ge fraction over B sites \\ \hline
        Sn\_frac & Sn fraction over B sites \\ \hline
        Pb\_frac & Pb fraction over B sites \\ \hline
        Cl\_frac & Cl fraction over X sites \\ \hline
        Br\_frac & Br fraction over X sites \\ \hline
        I\_frac & I fraction over X sites \\ \hline
        A\_ion\_rad & Ion radius of A site elements \\ \hline
        A\_BP & Boiling point of A site elements \\ \hline
        A\_MP & Melting point of A site elements \\ \hline
        A\_dens & Density of A site elements \\ \hline
        A\_at\_wt & Atomic weight of A site elements \\ \hline
        A\_EA & Electron affinity of A site elements \\ \hline
        A\_IE & Ionization energy of A site elements \\ \hline
        A\_hof & Heat of formation of A site elements \\ \hline
        A\_hov & Heat of vaporization of A site elements \\ \hline
        A\_En & Electronegativity of A site elements \\ \hline
        A\_at\_num & Atomic number of A site elements \\ \hline
        A\_period & Period number of of A site elements \\ \hline
        B\_ion\_rad & Ion radius of B site elements \\ \hline
        B\_BP & Boiling point of B site elements \\ \hline
        B\_MP & Melting point of B site elements \\ \hline
        B\_dens & Density of B site elements \\ \hline
        B\_at\_wt & Atomic weight of B site elements \\ \hline
        B\_EA & Electron affinity of B site elements \\ \hline
        B\_IE & Ionization energy of B site elements \\ \hline
        B\_hof & Heat of formation of B site elements \\ \hline
        B\_hov & Heat of vaporization of B site elements \\ \hline
        B\_En & Electronegativity of B site elements \\ \hline
        B\_at\_num & Atomic number of B site elements \\ \hline
        B\_period & Period number of of B site elements \\ \hline
        X\_ion\_rad & Ion radius of X site elements \\ \hline
        X\_BP & Boiling point of X site elements \\ \hline
        X\_MP & Melting point of X site elements \\ \hline
        X\_dens & Density of X site elements \\ \hline
        X\_at\_wt & Atomic weight of X site elements \\ \hline
        X\_EA & Electron affinity of X site elements \\ \hline
        X\_IE & Ionization energy of X site elements \\ \hline
        X\_hof & Heat of formation of X site elements \\ \hline
        X\_hov & Heat of vaporization of X site elements \\ \hline
        X\_En & Electronegativity of X site elements \\ \hline
        X\_at\_num & Atomic number of X site elements \\ \hline
        X\_period & Period number of of X site elements \\ \hline
    \end{tabular}
    \caption{\label{table:details_pearson} Expansion of each compositional and elemental property descriptor label used in this work.}
\end{table}

\begin{table}[!ht]
    \centering
    \begin{tabular}{|c|c|c|c|}
    \hline
        \textbf{Formula} & \textbf{PBE Band Gap (eV)} & \textbf{PBE Decomposition Energy (eV p.f.u.)} & \textbf{SLME at 5um thickness} \\ \hline
        $MAGeBr_3$ & 1.61 & -0.25 & 20.36\% \\ \hline
        $MASnCl_3$ & 1.58 & -0.25 & 16.53\% \\ \hline
        $MASnBr_3$ & 1.26 & -0.23 & 19.81\% \\ \hline
        $MAPbBr_3$ & 1.97 & -0.15 & 11.61\% \\ \hline
        $CsPbBr_3$ & 1.77 & -0.17 & 15.35\% \\ \hline
        $FAPbI_3$ & 1.94 & -0.44 & 18.64\% \\ \hline
        $K_{0.125}FA_{0.875}SnBr_3$ & 1.59 & -0.83 & 18.66\% \\ \hline
        $K_{0.25}Rb_{0.25}Cs_{0.125}MA_{0.375}PbBr_3$ & 2.11 & -0.07 & 10.26\% \\ \hline
        $Rb_{0.375}Cs_{0.25}MA_{0.375}PbBr_3$ & 2.03 & -0.08 & 12.24\% \\ \hline
        $FABa_{0.125}Pb_{0.875}I_3$ & 2.10 & -0.42 & 14.94\% \\ \hline
        $FABa_{0.25}Pb_{0.75}I_3$ & 2.35 & -0.46 & 12.03\% \\ \hline
        $FABa_{0.375}Pb_{0.625}I_3$ & 2.29 & -0.46 & 11.73\% \\ \hline
        $CsCa_{0.125}Ba_{0.125}Ge_{0.25}Sn_{0.5}Br_3$ & 1.50 & -0.10 & 21.85\% \\ \hline
        $FABa_{0.125}Ge_{0.25}Sn_{0.125}Pb_{0.5}I_3$ & 1.63 & -0.44 & 18.98\% \\ \hline
        $MACa_{0.125}Sn_{0.75}Pb_{0.125}I_3$ & 1.04 & -0.08 & 21.99\% \\ \hline
        $MABa_{0.25}Ge_{0.125}Sn_{0.625}I_3$ & 1.48 & 0.000 & 22.13\% \\ \hline
        $MACa_{0.125}Sn_{0.5}Pb_{0.375}Cl_3$ & 2.01 & -0.21 & 12.34\% \\ \hline
        $RbCa_{0.25}Ge_{0.25}Sn_{0.375}Pb_{0.125}Br_3$ & 1.54 & -0.01 & 22.17\% \\ \hline
        $FABa_{0.125}Ge_{0.125}Sn_{0.75}I_3$ & 1.26 & -0.55 & 21.85\% \\ \hline
        $MACa_{0.125}Ge_{0.375}Sn_{0.5}Cl_3$ & 2.04 & -0.25 & 13.14\% \\ \hline
        $MABa_{0.125}Pb_{0.875}Br_3$ & 2.24 & -0.12 & 11.21\% \\ \hline
        $RbGe_{0.875}Pb_{0.125}Cl_3$ & 1.09 & -0.24 & 18.15\% \\ \hline
        $FACa_{0.125}Pb_{0.875}I_3$ & 1.99 & -0.42 & 17.12\% \\ \hline
        $MABa_{0.125}Sn_{0.875}I_3$ & 1.11 & -0.08 & 23.15\% \\ \hline
        $MACa_{0.125}Pb_{0.875}Br_3$ & 2.14 & -0.13 & 11.13\% \\ \hline
        $MASr_{0.125}Pb_{0.875}Br_3$ & 2.17 & -0.13 & 10.99\% \\ \hline
        $MAGe_{0.125}Pb_{0.875}Br_3$ & 2.08 & -0.16 & 11.73\% \\ \hline
        $MASn_{0.125}Pb_{0.875}Br_3$ & 1.80 & -0.16 & 14.01\% \\ \hline
        $CsSnI_{1.875}Br_{1.125}$ & 2.47 & -0.13 & 10.92\% \\ \hline
        $CsPbI_{0.75}Br_{2.25}$ & 2.49 & -0.10 & 16.62\% \\ \hline
        $CsPbBr_{2.625}Cl_{0.375}$ & 1.98 & -0.16 & 13.62\% \\ \hline
        $MAPbI_{2.25}Br_{0.75}$ & 1.53 & -0.79 & 17.79\% \\ \hline
    \end{tabular}
    \caption{\label{table:PBE_screen_result} Chemical formulas and computed properties for 32 HaPs with desirable properties selected from the PBE dataset.}
\end{table}

\begin{table}[!ht]
    \centering
    \begin{tabular}{|c|c|c|c|}
    \hline
        \textbf{Formula} &  \textbf{Band Gap (eV)} &\textbf{Decomposition Energy (eV p.f.u.)} & \textbf{SLME at 5um thickness} \\ \hline
        $CsPbI_{0.75}Br_{2.25}$  & 1.24 & -0.52& 15.19\% \\ \hline
        $MACa_{0.125}Sn_{0.75}Pb_{0.125}I_3$  & 1.23 & -0.09& 15.12\% \\ \hline
        $CsPbBr_3$  & 1.42 & -0.55 & 15.05\% \\ \hline
        $MABa_{0.125}Sn_{0.875}I_3$  & 1.41 & -0.09 & 14.68\% \\ \hline
        $RbCa_{0.25}Ge_{0.25}Sn_{0.375}Pb_{0.125}Br_3$  & 1.88 & -0.12 & 13.75\% \\ \hline
        $MAPbBr_3$  & 1.72 & -0.39& 13.37\% \\ \hline
        $FABa_{0.25}Pb_{0.75}I_3$ & 2.23  & -1.22& 13.23\% \\ \hline
        $MASnBr_3$  & 1.69 & -0.31 & 13.22\% \\ \hline
        $CsCa_{0.125}Ba_{0.125}Ge_{0.25}Sn_{0.5}Br_3$  & 2.01 & -0.33 & 12.69\% \\ \hline
        $MABa_{0.125}Pb_{0.875}Br_3$  & 2.11 & -0.34 & 12.35\% \\ \hline
        $RbGe_{0.875}Pb_{0.125}Cl_3$ & 1.58 & -0.07 & 12.29\% \\ \hline
        $MABa_{0.25}Ge_{0.125}Sn_{0.625}I_3$  & 1.91 & -0.03 & 12.08\% \\ \hline
        $K_{0.25}Rb_{0.25}Cs_{0.125}MA_{0.375}PbBr_3$  & 1.99 & -0.60 & 11.27\% \\ \hline
        $MAGeBr_3$  & 2.25 & -0.34 & 10.67\% \\ \hline
    \end{tabular}
    \caption{\label{table:PBE_screen_result} Chemical formulas and computed properties for 14 HaPs with desirable properties selected from the HSE-PBE-SOC dataset.}
\end{table}

\end{document}